\documentclass[12pt,twoside]{article}
\newcommand{\refer}{\footnotesize \begin{thebibliography}{99}}
\newcommand{\erefer}{\end{thebibliography} \normalsize}
\newcommand{\res}{\setcounter{figure}{0} \setcounter{table}{0}%
\setcounter{equation}{0}}
\newcommand{\be}{\begin{equation}}
\newcommand{\ee}{\end{equation}}
\usepackage{epsf}
\begin{document}

\textwidth=6.5in
%\vspace*{-1cm}
%\begin{center}
%{\bf \Large FINAL DRAFT}
%\end{center}
\rightline{\today}

\begin{center}
{\Huge \bf Updated Report}\\
\vspace{0.2in}
{\Huge \bf Acceleration {\Large \bf of} Polarized Protons}\\
\vspace{0.2in}
\mbox{\Huge \bf {\Large \bf to} 120-150 GeV/c {\Large \bf at} Fermilab}\\
%\vspace{0.1in}
%{\Huge \bf at Fermilab}\\
\vspace{0.1in}
{\LARGE \rm SPIN@FERMI Collaboration}\\
\vspace{0.1in}
{\Large Michigan, Fermilab, Jefferson Lab,}\\
\vspace{0.1in}
{\Large Virginia, Argonne, Bonn, TRIUMF,}\\
\vspace{0.1in}
{\Large IHEP-Protvino, Novosibirsk}\\
\end{center}

\vspace{0.05in}
\addcontentsline{toc}{section}{Abstract}
%\begin{center}
%\large \bf ABSTRACT
%\end{center}
\normalsize
The SPIN@FERMI collaboration has updated its 1991-95 Reports on the acceleration of  
polarized protons in Fermilab's Main Injector, which was commissioned by Fermilab. 
This Updated Report summarizes some updated Physics Goals for a 120-150 GeV/c polarized proton beam. 
It also contains an updated discussion of the Modifications and Hardware needed for a 
polarized beam in the Main Injector, along with an updated Schedule and Budget. 
For reference, also attached to the e-mail containing this Updated Report are reprints 
of the 1992 and 1995 Reports on Polarized Beams at Fermilab by our SPIN collaboration.\\
\vspace*{0.2in}
Some highlights of the Update are: 
\vspace*{-3ex}
\begin{itemize}
\item
Two superconducting Siberian snakes in the Main Injector, one superconducting 60$^{\circ}$ 
rotator in the 120-150~GeV/c extraction line, a 4\% partial warm solenoidal Siberian snake 
in the 8.9~GeV/c Booster (oscillating with the Booster frequency) and some other minor hardware 
should allow about 75\% polarization to be maintained and manipulated in the RFQ, Linac, Booster, 
Recycler Ring and Main Injector, and then extracted to the experiments (See Fig. 1.9).
\item
Polarized ion sources now have intensities of 1.0 - 1.5 $m A$. 
With either the former IUCF Atomic Beam (ABS) type polarized ion source (which is now at 
Dubna), or the reconstructed and improved ZGS/AGS ABS, we expect to obtain an intensity of about 	
1 $m A$. With 10\% of the beam-time polarized, SeaQuest's  50 cm long $H_2$ target would have a 
time-averaged luminosity of about $2\cdot 10^{35}$ cm$^{-2}$ s$^{-1}$.  
\item
The estimated total cost of the project is about \$4 Million (2012 dollars). The 
construction time could be about 2 years after approval and funding.

\vspace*{-1ex}

\end{itemize}

\clearpage
\addcontentsline{toc}{section}{Contents}
\tableofcontents
%\addcontentsline{toc}{section}{LIST OF TABLES}
%\listoftable
\clearpage
%\cleardoublepage
%\addcontentsline{toc}{section}{LIST OF FIGURES}
%\listoffigure
%\newpage
%\setcounter{page}{1}
%\pagestyle{plain}
%\pagestyle{myheadings}
%\markboth{\protect\today}{}
%\pagenumbering{arabic}

\res
\setcounter{page}{3}
%\rightline {\today \qquad A. D. Krisch}
%\rightline {July 24, 1995 \qquad A. D. Krisch}

\section{Updated Report on Polarized Beams at Fermilab}

\subsection{SPIN@FERMI collaboration list}
\rightline{\today}
%\rightline{July 21, 2011}
%\noindent\normalsize\bf SPIN Collaboration \hfill \rm \today\\
\vspace*{.1in}

\noindent%\small 
E.~D.~Courant$^a$, A.~D.~Krisch, M.~A.~Leonova, A.~M.~T.~Lin, J.~Liu, \\
W.~Lorenzon, D.~A.~Nees, R.~S.~Raymond, D.~W.~Sivers$^b$, V.~K.~Wong\\ 
UNIVERSITY OF MICHIGAN, ANN ARBOR, U.S.A.
\vspace*{.12in}

\noindent%\small 
I.~Kourbanis \\
FERMILAB, BATAVIA, U.S.A.
\vspace*{.12in}

\noindent%\small 
Ya.~S.~Derbenev, V.~S.~Morozov \\
JEFFERSON LAB, NEWPORT NEWS, U.S.A.
\vspace*{.12in}

\noindent%\small 
D.~G.~Crabb \\
UNIVERSITY OF VIRGINIA, CHARLOTTESVILLE, U.S.A.
\vspace*{.12in}

\noindent%\small 
P.~E.~Reimer \\
ARGONNE NAT LAB, ARGONNE, U.S.A.
\vspace*{.12in}

\noindent%\small 
J.~R.~O'Fallon$^*$\\
DEPARTMENT OF ENERGY, WASHINGTON, U.S.A.
\vspace*{.12in}

\noindent%\small 
G.~Fidecaro$^*$, M.~Fidecaro$^*$ \\
CERN, GENEVA, SWITZERLAND.
\vspace*{.12in}

\noindent%\small 
F.~Hinterberger$^*$ \\
BONN UNIVERSITY, BONN, GERMANY.
\vspace*{.12in}

\noindent%\small 
S.~M.~Troshin, M.~N.~Ukhanov
 \\ 
INSTITUTE OF HIGH ENERGY PHYSICS,  PROTVINO, RUSSIA
\vspace*{.12in}

\noindent%\small 
A.~M.~Kondratenko \\
OOO "Zaryad", NOVOSIBIRSK, RUSSIA
\vspace*{.12in}

\noindent%\small 
W.~T.~H.~van~Oers \\ 
TRIUMF, VANCOUVER, CANADA
\vspace*{.12in}

%\footnotesize
\noindent The spokesperson for the SPIN@FERMI Collaboration is:\\[0.2ex]
\vspace{.2in}
\begin{tabular}{ll}
A. D. Krisch & Telephone: 734-936-1027\\
Randall Laboratory of Physics & Telefax: 734-936-0794\\
University of Michigan & E-mail: krisch@umich.edu\\
Ann Arbor, Michigan 48109-1040 USA&  \\
\end{tabular} 

\noindent $^{*}$ retired \\
\noindent Permanent address:\\
\begin{tabular}{llll} 
$a\,$ NYC\ &  &  &  \\ 
$b\,$ Portland\ &  &  &  \\
%$c\,$ \ &  &  &  \\

\end{tabular}

\normalsize
\pagebreak

\subsection{Experimental Overview}
\vspace{-0.3in}
\rightline{\small M.A.~Leonova, A.D.~Krisch, \normalsize}
\vspace{0.1in}
\bf Introduction. \rm
The interest in spin phenomena has significantly increased in recent years.  
It is now clear that  spin effects in high energy interactions provide 
essential information about the elementary particles' properties and structure. 
Recently, there has been significant progress in understanding the nucleon's 
longitudinal and transverse spin structure due to many polarization experiments 
done at SLC, HERA, CERN and RHIC. 
The Main Injector polarized proton beam would allow unique studies of spin phenomena 
such as the 1-spin asymmetry in all inclusive processes, including Drell-Yan 
and hadron and hyperon production. 
It would also allow both 1-spin and 2-spin asymmetry measurements of exclusive processes 
such as proton-proton elastic scattering at large $P_\perp^2$.
Thus, the Main Injector's very high intensity could test the validity of strong 
interaction theories at the far larger $P^2_{\perp}$ values possible at 120-150~GeV/c.    

\bf Polarized Drell-Yan Experiments. \rm
The E-866$^{\scriptsize \cite{moss} \rm}$ and E-906 (SeaQuest) collaborations 
have had a long-term interest in studying
Drell-Yan processes with a 120-150~GeV/c polarized beam. Details are given in Section 1.4.

\bf Polarized large $\mathbf{P_{\perp}^2}$ elastic and inclusive scattering. \rm 
Transverse spin effects appear experimentally to increase at large-$P_{\perp}$. 
A high intensity polarized beam could determine if these unexpected spin effects 
persist at the larger $P_{\perp}$ possible at the 120-150 GeV/c Main Injector. 
The SPIN@FERMI Collaboration hopes to continue studying the proton's transverse spin
structure by scattering a 120-150~GeV/c extracted polarized proton beam from a solid 
polarized proton target and a liquid hydrogen target. 
As shown in Fig.~1.1, a large left-right asymmetry $A_N$ was found in 
polarized proton-proton elastic scattering at large $P_\perp^2$$^{\scriptsize \cite{ref1.1} \rm}$.
Currently, the nucleon's transverse spin structure is unexplored experimentally above 
about $P_\perp^2 = 7$~(GeV/c)$^2$.
\vspace{-0.15in}
\begin{figure}[h!]
\centerline
{
\epsfysize=3.8in \epsfbox{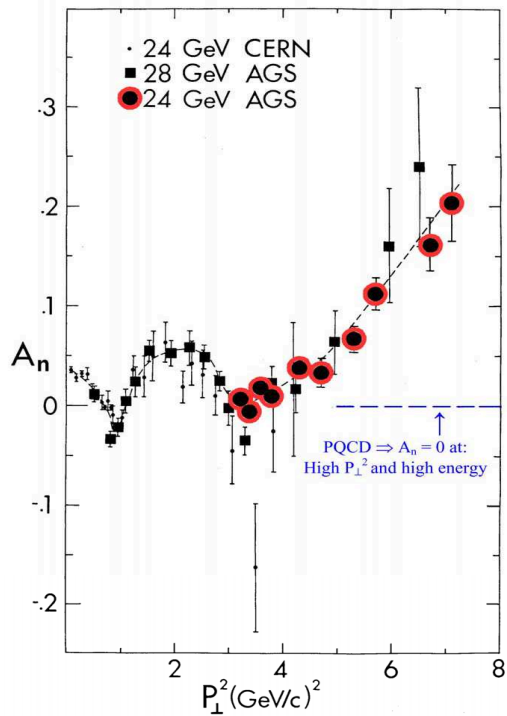}
}
\vspace{-0.15in}
\caption{\footnotesize Plot of $A_N$ against  
$P_{\perp}^{2}$ for proton-proton elastic scattering 
($p_{\uparrow} + p \rightarrow p + p$)$^{\scriptsize \cite{ref1.1} \rm}$. \normalsize}
\label{AP}
\end{figure}

\newpage

\begin{figure}[h!]
\vspace{-0.15in}
\centerline
{
\epsfysize=1.9in \epsfbox{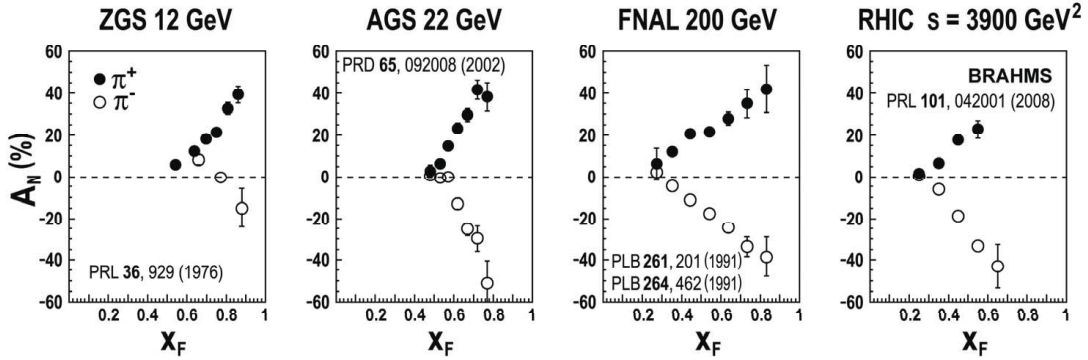}
}
\vspace{-0.2in}
\caption{\footnotesize Inclusive 1-spin $\pi^{+}$ and $\pi^{-}$ 
$A_N$ (left-right asymmetry) plotted 
against $X_{F}$$^{\scriptsize \cite{ref1.2} \rm}$. \normalsize}
\label{LP}
\end{figure}

Similar large asymmetries were found in large-$X_{F}$ inclusive pion 
production$^{\scriptsize \cite{ref1.2} \rm}$ 
from $P_{lab} = 12$~GeV/c to $s = 3900$~(GeV/c)$^2$, as shown in Fig.~\ref{LP}. 

There are 2 independent 1-spin $A_N$ asymmetries in large
$P^2_{\perp}$ elastic scattering (polarized beam and polarized target), 
\vspace{-0.15in}
\be
p_{\uparrow} + p \rightarrow p + p  
\hspace{0.2in} \mbox{and} \hspace{0.2in} 
p + p_{\uparrow} \rightarrow p + p. 
\vspace{-0.15in}
\ee
For identical particles, such as 2 protons, the 2 independent $A_N$ asymmetries must be equal.
These would be measured simultaneously with the 2-spin $A_{NN}$ asymmetry,
\vspace{-0.15in}
\be
p_{\uparrow}+p_{\uparrow} \rightarrow p+p.  
\vspace{-0.in}
\ee

\begin{figure}[h!]
\vspace{-0.2in}
\centerline
{
\epsfysize=4.1in \epsfbox{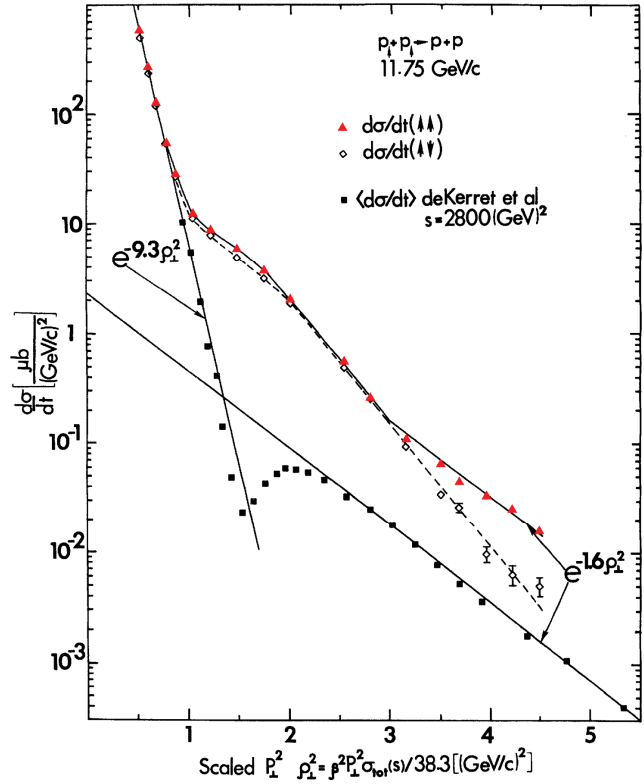}
}
\vspace{-0.1in}
\caption{\footnotesize Spin-spin effects in 2-spin proton-proton elastic scattering at large 
$P^2_{\perp}$$^{\scriptsize \cite{ref1.3} \rm}$. \normalsize}
\label{pol-cros-sec-elast}
\end{figure}

\vspace{-0.15in}
As shown in Fig.~1.3, a large and unexpected 2-spin asymmetry was
found at large $P^2_{\perp}$ near 12~GeV/c. One could determine if the large and still
unexplained $A_{NN}$ disappears or persists at the large $P^2_{\perp}$ available at the 
high-intensity 120-150~GeV/c Main Injector.

\begin{figure}[h!]
\centerline
{
\epsfysize=7.3in \epsfbox{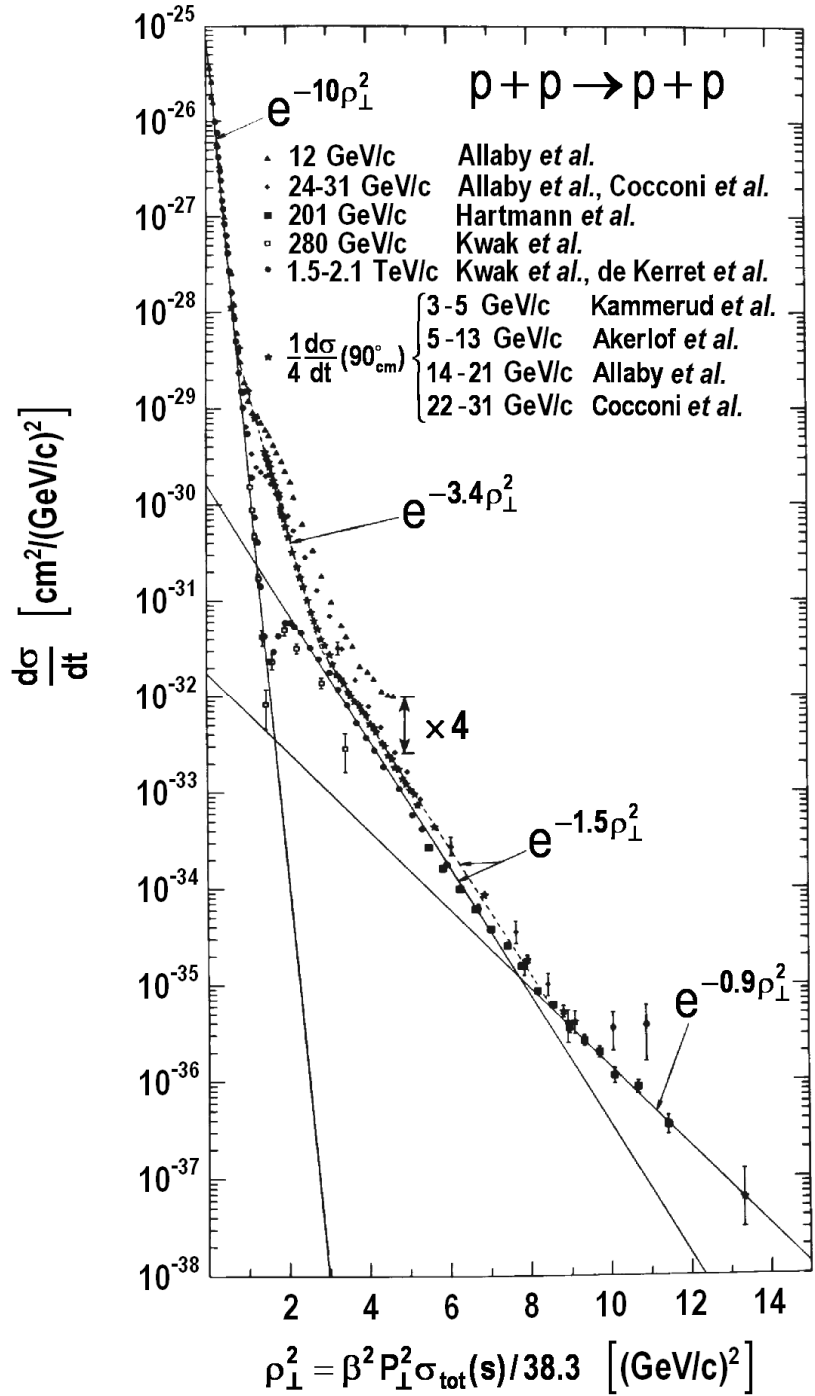}
}
\caption{\footnotesize All unpolarized elastic proton-proton cross section data above 3~GeV/c 
plotted against the scaled $P^2_{\perp}$ variable$^{\scriptsize \cite{fig4} \rm}$. \normalsize}
\label{unpol-crossec}
\end{figure}

Moreover, with the high intensity Main Injector, one could simultaneously measure the unpolarized 
proton-proton elastic cross section at large $P_\perp^2$ with much better precision than now 
exists. Figures 1.1, 1.3,~\ref{unpol-crossec} (and 1.6) show compilations of all 
existing data on the proton-proton elastic scattering's cross section and its $A_{N}$ and $A_{NN}$ asymmetries 
above a few GeV/c.

\pagebreak
\subsection{Theoretical Overview}
\vspace{-0.3in}
\rightline{\small D.W.~Sivers, M.A.Leonova, A.D. Krisch\normalsize}

Spin has been the quantum number that has mystified physicists since its publication by 
Uhlenbeck and Goudsmit in 1925$^{\scriptsize \cite{UG} \rm}$. 
Indeed, at the 1982 SPIN Symposium, C.N.~Yang stated 
"I for one suspect that the spin and general relativity are deeply entangled in a subtle way that 
we do not understand"$^{\scriptsize \cite{yang} \rm}$. 
The modern spin era began with Wolfenstein's$^{\scriptsize \cite{wolf} \rm}$ 
efforts to develop a formalism to describe spin experiments, followed by the discovery 
by Oxley {\it et al.} of large spin effects at 200~MeV$^{\scriptsize \cite{oxley} \rm}$. 
Fermi's last paper$^{\scriptsize \cite{fermi} \rm}$ 
focused on his amazement that the proton's spin, which had so little energy, was so important at 
200~MeV. This paper resulted in his ex-student Chamberlain and others starting a series of double 
and triple scattering experiments and developing polarized proton targets and using them 
with unpolarized beams. Next, the development of the 12 GeV/c ZGS polarized proton beam allowed 
many precise spin experiments, such as proton-proton elastic 
scattering experiments$^{\scriptsize \cite{ref1.3,crosbie} \rm}$, 
which found unexpectedly large $A_{N}$ and $A_{NN}$ asymmetries at high $P^2_{\perp}$. This started 
the era of GeV polarized beams and polarized targets at many high energy and nuclear accelerators and colliders and many theoretical efforts to try to understand the resulting data.

There was a belief that quantum chromodynamics (QCD) predicted that all transverse single spin 
asymmetries would vanish at large transverse momentum. This misconception can be traced to 
statements found in the paper of Kane, Pumplin and Repko$^{\scriptsize \cite{KPR} \rm}$.  
It correctly pointed out that large transverse single spin asymmetries are not 
generated in perturbative processes involving light quarks, 
\vspace{-0.13in}
\be
A_N \, d \sigma (q q_{\uparrow} \rightarrow q q) / d\sigma (q q \rightarrow q q) = 
\frac{\alpha_s(Q^2)}{Q} m_q f(\theta_{CM}). 
\label{perturb-proc-quarks}
\ee
But they mistakenly neglected other possible twist-3 mechanisms in a collinear factorization 
formulation of a hard scattering model, and used this to suggest the vanishing of 1-spin 
observables, such as $A_{N}$. During 1978-1988 this conclusion was widely accepted. 

The correct interpretation$^{\scriptsize \cite{sivers1} \rm}$ of their result was that, 
since perturbative processes involving light quarks do not themselves 
generate 1-spin asymmetries, they could be used to probe the asymmetries 
caused by the soft nonperturbative dynamics of QCD due to the interplay 
of confinement and dynamic chiral symmetry breaking. 
The asymmetries generated by such spin-orbit dynamics can be parameterized into 
$k_T$-dependent distribution functions (Sivers functions$^{\scriptsize \cite{sivers1} \rm}$ 
or Boer-Mulders distributions$^{\scriptsize \cite{BM-dis} \rm}$) 
or into $k_T$-dependent fragmentation functions (Collins 
functions$^{\scriptsize \cite{collins1} \rm}$ or polarizing fragmentation 
functions$^{\scriptsize \cite{sivers1,ff} \rm}$).  
They can also be parameterized into specific twist-3 operators in a collinear 
factorization approach. Mulders and his collaborators$^{\scriptsize \cite{ff} \rm}$ 
classified the appropriate operators for 1-spin asymmetries but, mistakenly, 
called them T-odd suggesting that they violated time reversal invariance.  
In truth, the symmetry they violated involves a transformation related to the Hodge 
dual operator of differential geometry$^{\scriptsize \cite{sivers2} \rm}$. 
The Trento Conventions for transverse spin asymmetries are described 
in Ref$^{\scriptsize \cite{trento} \rm}$. The subject received a boost 
when Heppelmann, Collins and Ladinsky$^{\scriptsize \cite{CHL} \rm}$ 
noted that the quark transversity distributions, $\delta^T q(x)$, defined 
by Ralston and Soper$^{\scriptsize \cite{RS} \rm}$ and renamed 
by Jaffe and Ji$^{\scriptsize \cite{JJ} \rm}$, 
could be measured in semi-inclusive deep inelastic scattering (SIDIS),
\vspace{-0.05in}
\be
A_N \, d \sigma (l p_{\uparrow} \rightarrow l^{\prime} \pi X) \propto 
\delta^T q(x) \otimes H_1(z).
\label{semi-inclus-DIS}
\ee
Here $H_1 (z)$ is the Collins function$^{\scriptsize \cite{collins1} \rm}$ that 
defines an asymmetry in the fragmentation of a transversely polarized quark.  
Asymmetries involving fragmentation functions can be separated from those involving 
distributions in SIDIS and in the Drell Yan$^{\scriptsize \cite{DY} \rm}$ process.  
In hadron-hadron collisions they can also be separated at the level of two-particle 
correlations in the final state.  
A comprehensive phenomenological fit to asymmetries in $e^{+} + e^{-} \rightarrow hadrons$, 
semi-inclusive DIS and inclusive production in polarized hadron-hadron scattering, 
has been published by the Turin group$^{\scriptsize \cite{Anselmino} \rm}$.  
They fit the transversity distributions for up and down quarks, favored and disfavored 
Collins functions, and orbital distributions for up and down quarks.  
This phenomenology is currently being extended to NLO in QCD perturbation theory.  
An important feature of ''T-odd'' distribution functions is that they are required to display 
a dramatic process dependence in order to be consistent with a gauge formulation of QCD.  
This result can be called Collins conjugation$^{\scriptsize \cite{collins2} \rm}$.  
It needs to be tested.  
One comparison involves the measurement of orbital distributions (which are called 
Sivers functions) in DIS and in the DY process,
\vspace{-0.1in}
\be
f^{\perp q}_{1T}(DIS) = - f^{\perp q}_{1T}(DY).
\label{DY-orbital-DIS}
\ee
However, others can also be considered.  In particular, the Boer-Mulders distributions 
exhibit the same process dependence as orbital distributions.  
Related processes involving associated baryon production can also be studied..

Exclusive processes in QCD involve local descriptions in terms of, so called, 
generalized parton distributions (GPD's) and effective field theories incorporating 
constraints generated by crossing and analyticity.  
Quantum chromodynamics predicts that elastic transverse spin asymmetries for 
hadron-hadron scattering at large transverse momenta involve a combination of the 
Chou-Yang$^{\scriptsize \cite{CY} \rm}$ mechanism involving orbital angular momentum 
and the Brodsky-Lepage$^{\scriptsize \cite{LB} \rm}$ effective field theory 
which involves truncation of the Fock states combined with power-law approximations 
to effective form factors. 
However, the data shown in Figs. 1.1, 1.2 and 1.3 are too small a set of data to 
conclusively filter the various theoretical approaches. 
Its range needs to be increased. High intensity 120-150 GeV/c polarized proton beams from the 
Main Injector could allow a comprehensive experimental program of transverse single-spin and 
double-spin experiments.

\vspace{0.05in}
\noindent \bf Single-Spin Asymmetries \rm
\vspace{0.05in}

\noindent
These studies take advantage of the high luminosity possible with a polarized beam scattering on 
an unpolarized target, together with the flexibility of incorporating high-resolution 
measurements of momentum with particle identification in fixed-target experiments.  
A list of important experiments would include:

\vspace{0.05in}
\noindent
\bf 1. Polarized Drell Yan asymmetries \rm
\vspace{-0.1in}
\be
A_N \, d \sigma (p_{\uparrow} + p \rightarrow \mu^{+} + \mu^{-} + anything)
\label{pol-DY}
\ee
This is a fundamental measurement that can be used to test the validity of the gauge 
formulation of QCD in regions where the fundamental degrees of freedom cannot be 
clearly isolated.  In this sense, Collins conjugation can be formulated in analogy 
to the Bohm-Aharanov test of the gauge formulation of QED.

\newpage
\vspace{0.0in}
\noindent
\bf 2. Spin asymmetries in Baryon production \rm
\vspace{-0.1in}
\be
A_N \, d \sigma (p_{\uparrow} + p \rightarrow B + anything)
\label{tgt-baryon-prod}
\ee
These asymmetries involve mechanisms closely related to those responsible for the production 
of baryons with polarization ($P$) from unpolarized scattering processes.
\be
P d \sigma (p + p \rightarrow B_{\uparrow} + anything)
\label{unpol-baryon-prod}
\ee
Inclusive hyperon polarization [Eq.~(1.8)] was studied experimentally in the late 1970s at 
Fermilab$^{\scriptsize \cite{hyp-incl} \rm}$.

\vspace{0.15in}
\noindent
\bf 3.  Elastic Scattering Single Spin Asymmetries \rm
\vspace{-0.1in}
\be
A_N \, d \sigma_{el} (p_{\uparrow} + p \rightarrow p + p)
\label{AN-elast}
\ee
This could test the combination of the Brodsky-Lepage$^{\scriptsize \cite{LB} \rm}$ effective 
field theory with the Chou-Yang$^{\scriptsize \cite{CY} \rm}$ formulation of elastic 
scattering involving orbiting constituents. 
It provides an independent measurement of the mean orbital angular momentum of the rotating charges.

\vspace{0.15in}
\noindent
\bf 4. Spin asymmetries in inclusive pseudoscalar and vector meson production \rm
\vspace{-0.1in}
\begin{eqnarray}
A_N \, d \sigma (p_{\uparrow} + p \rightarrow M + anything), \\
{\mbox{where~~}} M = \pi, K, \eta, \eta^{\prime}, \rho, K^{\star}, \omega, \phi \nonumber.
\label{AN-meson}
\end{eqnarray}
These processes were studied experimentally in the late 1980s at 
Fermilab$^{\scriptsize \cite{E904} \rm}$.
Much higher intensity studies could provide a more precise understanding of Collins functions 
and precision measurements of orbital distributions.  

\vspace{0.15in}
\noindent
\bf 5. Two-particle correlations \rm 
\vspace{-0.1in}
\be
A_N \, d \sigma (p_{\uparrow} + p \rightarrow \phi^{+} +  \phi^{-} + anything)
\label{two-ptcl-correl}
\ee 
Non-resonant two-particle correlations can be used to distinguish between asymmetries that 
occur in the fragmentation process (Collins functions) from those that occur in the proton's 
distribution orbital distributions (Sivers functions).  These experiments take advantage of the 
ability to combine accurate momentum measurements with particle identification in fixed 
target experiments.

\vspace{0.15in}
\noindent
\bf 6. Spin asymmetries in inclusive $J / \psi, \, \psi^{\prime}$ and Charm production \rm
\vspace{-0.1in}
\be
A_N \, d \sigma (p_{\uparrow} + p \rightarrow J / \psi + anything); \,
A_N \, d \sigma (p_{\uparrow} + p \rightarrow \psi^{\prime} + anything) 
\label{tgt-AN-psi}
\ee

\noindent Like the Drell Yan process, these processes are free from important fragmentation asymmetries 
and can be used to measure gluon orbital distribution functions and, hence, gluon 
orbital angular momentum.  Again, the access to forward kinematics is an immense 
advantage of fixed target kinematics.

\newpage
\vspace{0.05in}
\noindent
\bf Double-Spin Asymmetries \rm 
\vspace{0.05in}

\noindent
Combining a polarized beam with a polarized target provides access to two spin asymmetries.  
Here we only mention a few transverse $A_{NN}$ asymmetries, which could be studied 
experimentally using a Main Injector polarized beam:

\vspace{0.05in}
\noindent
\bf 1. Drell-Yan 2-Spin Experiment\rm
\vspace{-0.1in}
\be
\vspace{-0.05in}
A_{NN} \, d \sigma (p_{\uparrow} + p_{\uparrow} \rightarrow \mu^{+} + \mu^{-} + anything)
\label{DY-ANN}
\ee
There is a strong prejudice that antiquark transversities are very small, but this belief 
needs to be confronted with experiment.  In addition, there are many other asymmetries 
involving different angular distributions that appear with the extra degree of freedom.

\vspace{0.15in}
\noindent
\bf 2.  High-$\mathbf{P_{\perp}^2}$ 2-Spin proton-proton elastic Experiment \rm
\vspace{-0.1in}
\be
A_{NN} \, d \sigma_{elastic} (p_{\uparrow} + p_{\uparrow} \rightarrow p + p)
\label{elast-ANN}
\ee
Extending these experiments$^{\scriptsize \cite{ref1.3,crosbie} \rm}$ to 
the $P_{\perp}^2$ available at the Main Injector would provide a tremendous 
expansion of the range in which the Brodsky-Lepage$^{\scriptsize \cite{LB} \rm}$ 
approach to exclusive processes has been measured.

\pagebreak
\subsection{Physics with 120-150 GeV/c polarized beams}
\vspace{-0.1in}
\rightline{\small  W.~Lorenzon, A.D.~Krisch \normalsize}
\vspace{0.1in}

A 120-150~GeV polarized beam at the Fermilab Main Injector and a liquid 
hydrogen target and/or a solid polarized proton target (PPT) could allow a wide
range of 1-spin and 2-spin asymmetry measurements. Particularly interesting would be: 
Drell-Yan$^{\scriptsize \cite{DY} \rm}$ scattering experiments with a transversely 
polarized protons beam on unpolarized liquid hydrogen targets; 
and large-$P_{\perp}^2$ elastic scattering in view of the still unexplained huge 
transverse spin-effects at 12 to 28~GeV/c found at the ZGS$^{\scriptsize \cite{ZGS} \rm}$ 
and AGS$^{\scriptsize \cite{AGS} \rm}$.
Some transversely polarized hadron measurements include:
{\protect\vspace*{-1.5ex}}
\begin{itemize}
\item
the Sivers asymmetries in high precision polarized Drell-Yan experiments;
\vspace*{-2ex}
\item
the 1-spin $A_N$ and 2-spin $A_{NN}$ in large-$P_\perp^2$~ proton-proton elastic scattering;
\vspace*{-2ex}
\item
the 2-spin proton-proton total cross section ${\sigma_{tot}A_{NN}}$;
\vspace*{-2ex}
\item
the 2-spin $D_{NN}$ of $\Lambda$-hyperon polarization via its self-analyzing decay;
\vspace*{-2ex} 
\item
the left-right asymmetry in $\Sigma^0$-production or $\rho$-production; 
\vspace*{-2ex}
\item
the left-right asymmetries in inclusive pion and kaon production.
\vspace*{-1ex}
\end{itemize}

\vspace{0.005in}
\noindent
\bf 1. Polarized Drell-Yan scattering \rm has become a major milestone in the 
hadronic physics community, motivated by a fundamental prediction of QCD that
postulates a sign change in the Sivers function$^{\scriptsize \cite{sivers1} \rm}$ 
measured in Drell-Yan scattering as compared to semi-inclusive deep inelastic 
scattering (SIDIS)$^{\scriptsize \cite{collins2,Brodsky2002} \rm}$.
Each quark and antiquark flavor has its own Sivers function described by a 
transverse-momentum dependent distribution function that captures non-perturbative 
spin-orbit effects inside a polarized proton. 
The experimental verification of the sign change goes to the heart of the gauge 
formulation of QCD and would fundamentally test the factorization approach to the 
description of processes sensitive to transverse parton momenta. It would be crucial 
to confirm the validity of our present conceptual framework for analyzing hard 
hadronic reactions.

\begin{center}
\vspace{-0.1in}
\begin{figure}[htb]
\centerline
{
\epsfysize=3.0in \epsfbox{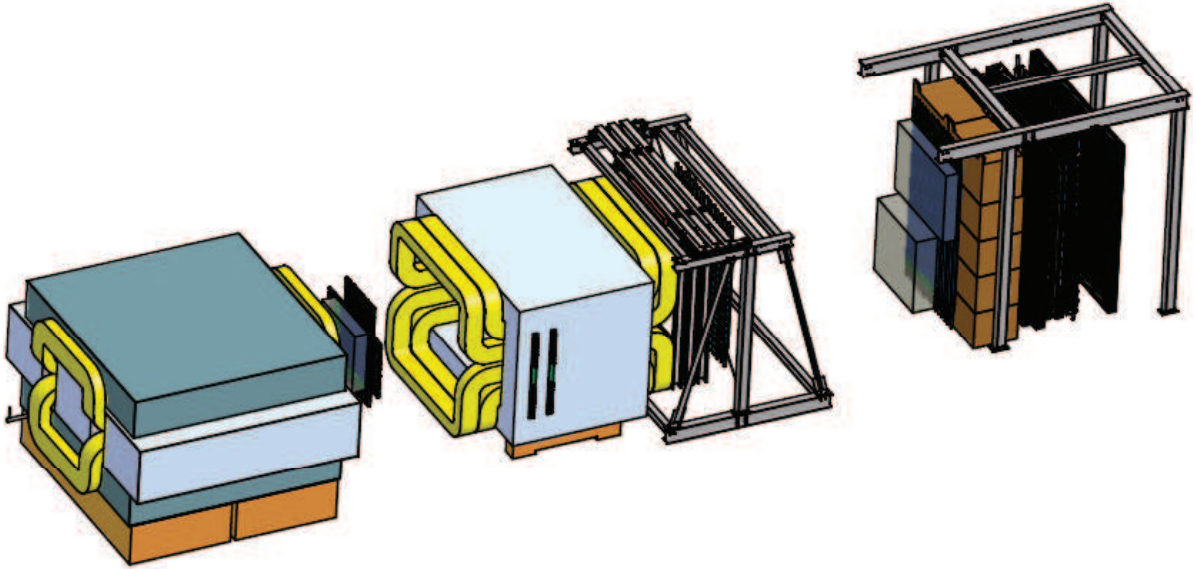}
}
\vspace{-0.05in}
\caption{\footnotesize Schematic layout of the polarized Drell-Yan spectrometer.
The polarized beam enters from the left and hits a 50-cm long unpolarized
target before it is stopped in the 5-m long solid-iron magnet. \normalsize}
\label{fig:SeaQuest-spectrometer}
\end{figure}
\end{center}

The HERMES$^{\scriptsize \cite{hermes2005} \rm}$ and 
COMPASS$^{\scriptsize \cite{compass2005} \rm}$ experiments have 
measured single transverse spin asymmetries and performed global fits to the 
Sivers asymmetries with high precision. 
In order to make a meaningful comparison of shape and sign, comparable measurements 
are needed for single spin asymmetries in the Drell-Yan process. 
While many experiments around the globe aim to measure polarized Drell-Yan either 
with a polarized beam or a polarized target, none of them is optimized for Drell-Yan 
except for the SeaQuest di-muon spectrometer at the Fermilab Main Injector. 
SeaQuest will use 5-s long spills of $2\cdot10^{12}$ protons/s each minute 
(I$_{\rm av}=1.6\cdot10^{11}$ protons/s) on a 50-cm long liquid hydrogen 
(or deuterium) target ($N_p = 2.1\cdot10^{24}$ cm$^{-2}$). 
This results in an average 
luminosity of $3.4\cdot10^{35}$~cm$^{-2}$s$^{-1}$ and a total integrated 
beam of $3.4\cdot10^{18}$ protons on target over a period of 2 to 3 years of running.

The big attraction for a polarized Drell-Yan program at the Fermilab Main
Injector is a spectrometer and hydrogen target that are well-understood,
fully functioning, and optimized for Drell-Yan at the end of data collection
for the SeaQuest experiment, shown in Fig.~\ref{fig:SeaQuest-spectrometer}.
Based on the study presented in this report and experience from current
polarized ion sources, it is expected that an ion source that produces
1$\;$mA at the source can deliver up to 150$\;n$A (about $1\cdot10^{12}$ p/s) to
the experiment by using 30 two-second cycles and slip stacking into the Main
Injector. Assuming that 50\% of the total beam time is allocated to the
experiment, a luminosity of $1\cdot10^{36}$~cm$^{-2}$s$^{-1}$ can be
obtained. It is important to note that even if only 10\% of the
available beam time was allocated to the experiment, a luminosity of
$2.0\cdot10^{35}$~cm$^{-2}$s$^{-1}$ is still very competitive. In
addition, the SeaQuest spectrometer accommodates a large coverage in $x$,
i.e., $x_1 = 0.3 -0.9$ covering the valence quark region, and $x_2 = 0.1-0.5$
covering the sea quark region. While the Sivers function can be measured for
both the valence quarks or the sea quarks, sea quark effects might be small
due to competing processes, while valence quark effects are generally
expected to be large$^{\scriptsize \cite{VQL} \rm}$. 
Thus, using a polarized beam might have a substantial
advantage over a polarized target. The combination of high luminosity, large
$x$-coverage and a high-intensity polarized beam makes Fermilab arguably the
best place to measure Drell-Yan scattering with high precision.

\begin{figure}[p]
\centerline
{
\epsfysize=3.95in \epsfbox{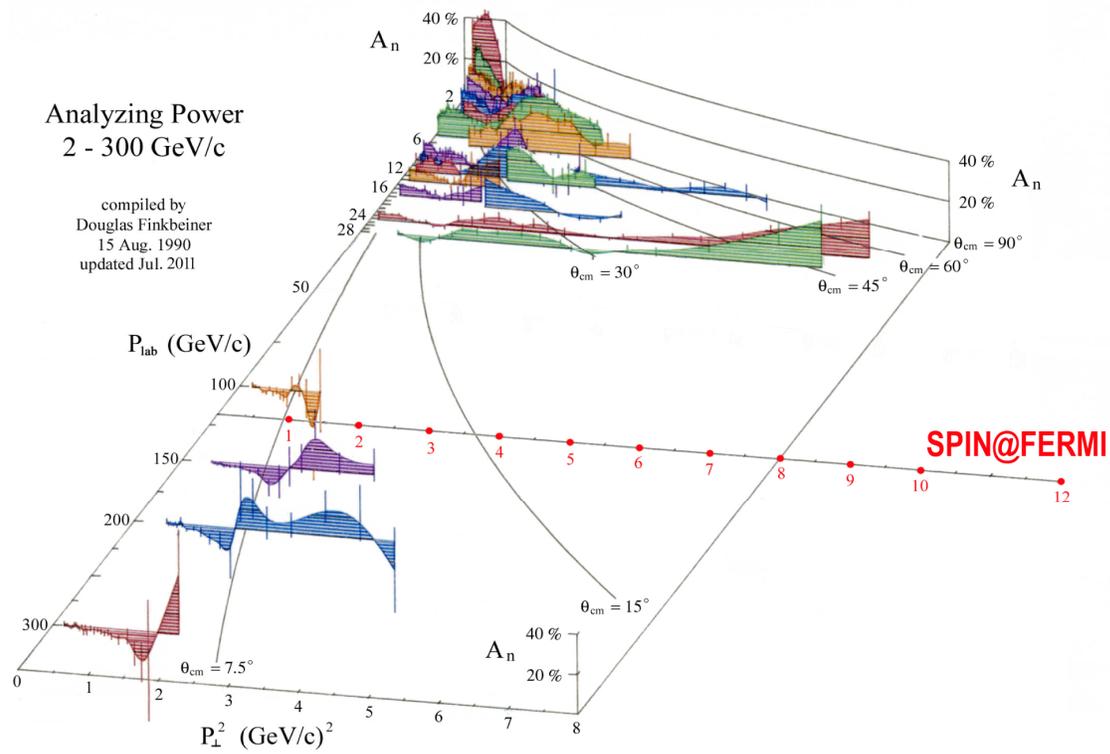}
}
\caption{\footnotesize Compilation of all elastic $A_N$ data above 2 GeV/c showing range of possible SPIN@FERMI experiment. \normalsize}
\label{an-compilation}
\end{figure}

\begin{figure}[p]
\centerline
{
\epsfysize=3.95in \epsfbox{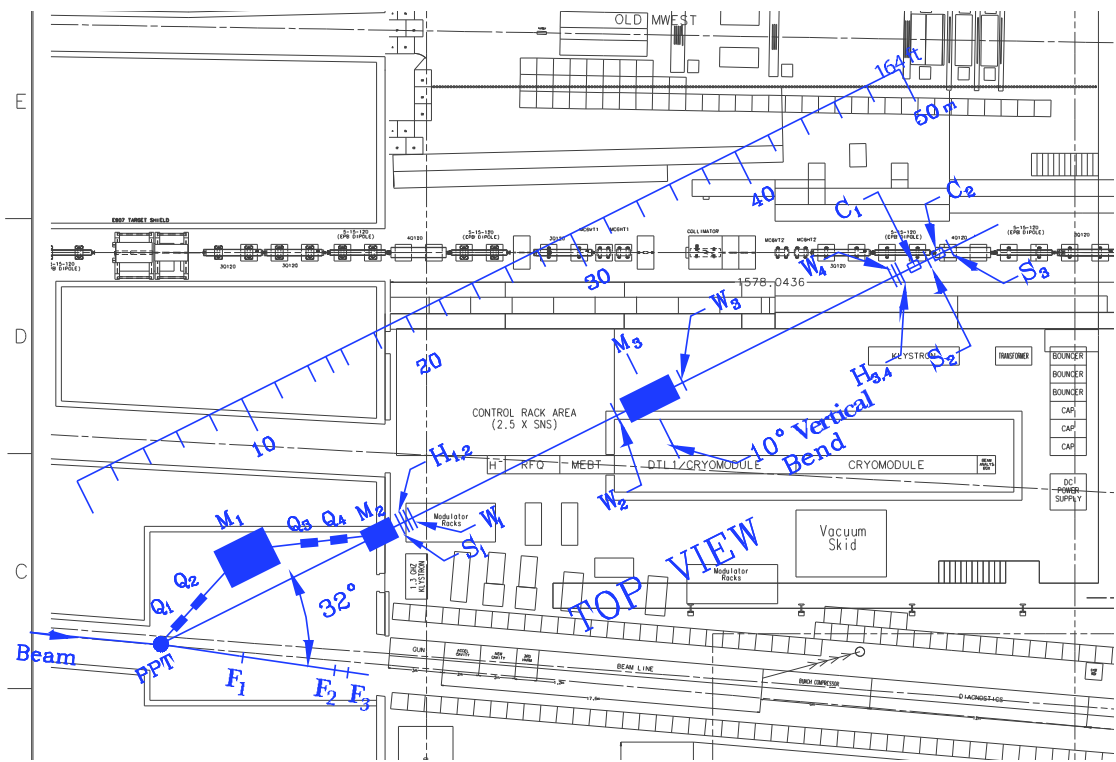}
}
\caption{\footnotesize Possible SPIN@FERMI experiment layout in the Meson Hall. \normalsize}
\label{spectrom}
\end{figure}

\vspace{0.005in}
\noindent
\bf 2. Polarized elastic scattering \rm could shed new light on 
the nature of the strong interaction. As discussed in the attached March 1992 
Polarized Main Injector Report, one could do many fixed-target elastic 
polarized beam experiments with very high luminosity. These 120-150~GeV/c 
experiments could use the existing Michigan solid Polarized Proton Target (PPT) 
which operated at the AGS with a time-averaged beam intensity of over 
\mbox{$10^{11}$ protons/s} and polarization of about 85\%.  
With a 10\% time share available for polarized beam, the expected 120-150~GeV/c 
polarized beam intensity of $\sim$ $1\cdot 10^{12}$ protons per Main Injector cycle, 
would give an average beam intensity of $\sim$ \mbox{$1 \cdot 10^{11}$ protons/s} 
scattering from this solid polarized target ($N_p = 2 \cdot 10^{23}$ cm$^{-2}$); 
the proton-proton luminosity would be about 
\vspace{-0.1in}
\be 
{\cal L}=2\cdot10^{34}~cm^{-2}~s^{-1}\ . 
\ee
A high quality recoil proton spectrometer could simultaneously extend the range 
of $A_N$, $A_{NN}$ and the elastic cross section in the unexplored large-$P_\perp^2$ 
region shown in Figs.~1.1, 1.3 and~\ref{an-compilation}, and Fig.~1.4, respectively.
A possible placement for the solid PPT and a large-$P^2_\perp$
elastic recoil spectrometer in the Meson Hall is shown in Fig.~\ref{spectrom}. 
Moreover, the experiment was originally proposed for the 400 GeV/c UNK 
where the non-elastic background was considerably larger; thus, the recoil 
spectrometer might be significantly shortened to fit better into the 
Meson Hall, as was done in the U-70 Hall, after UNK was suspended.
These experiments could run simultaneously with the Main Injector 
running in the polarized and unpolarized mode on interspersed pulses.  
These 120-150~GeV/c fixed-target spin experiments may provide further
justification for developing polarization capability at Fermilab. 
The SPIN@FERMI collaboration has been very interested in 120-150~GeV/c large-P$_{\perp}^2$ 
elastic proton-proton spin experiments since the 1980s.


The luminosity of $\sim$$2\cdot 10^{34}$ cm$^{-2}$ s$^{-1}$ should be adequate 
for elastic scattering out to $P_{\perp}^{2}$ of $\sim$12~(GeV/c)$^{2}$.
The expected polarized elastic event rates per day are about:
\vspace*{-0.1in}
\begin{tabbing} 
\hspace*{1.95in} $200000$ at $P_\perp^2$ \= = 2 (GeV/c)$^2$; \\
\hspace*{2.05in} $40000$ at $P_\perp^2$ \> = 4 (GeV/c)$^2$; \\
\hspace*{2.15in} $4100$ at $P_\perp^2$ \> = 6 (GeV/c)$^2$; \\ 
\hspace*{2.2in} $480$ at $P_\perp^2$ \> = 8 (GeV/c)$^2$; \\ 
\hspace*{2.3in} $80$ at $P_\perp^2$ \> = 10 (GeV/c)$^2$; \\ 
\hspace*{2.3in} $20$ at $P_\perp^2$ \> = 12 (GeV/c)$^2$. \\
\end{tabbing}
\vspace*{-0.3in}
\noindent We might later increase these event rates by:
\vspace*{-1ex}
\begin{itemize}
\item increasing the polarized ion source intensity above 1.5~mA; 
\vspace*{-2ex}
\item further improving the PPT for running with high beam intensity.
\end{itemize}

\vspace{0.005in}
\noindent
\bf 3. Polarized Large-$\mathbf{P_{\perp}}$ inclusive processes. 
\rm Measuring inclusive spin effects at the high-intensity Main Injector 
could provide a precise new probe of the strong interaction at very large $P_\perp$.  
One could precisely measure, at very large $P_\perp$, the 1-spin transverse asymmetries 
($A_N$) in inclusive processes such as: 
\be
p_{\uparrow}+p \rightarrow \pi^{\pm} + anything, 
\ee
\be
p_{\uparrow}+p \rightarrow K^{\pm} + anything. 
\ee

\begin{figure}[htb]
%\vspace{0.1in}
\centerline
{
\epsfysize=4.0in \epsfbox{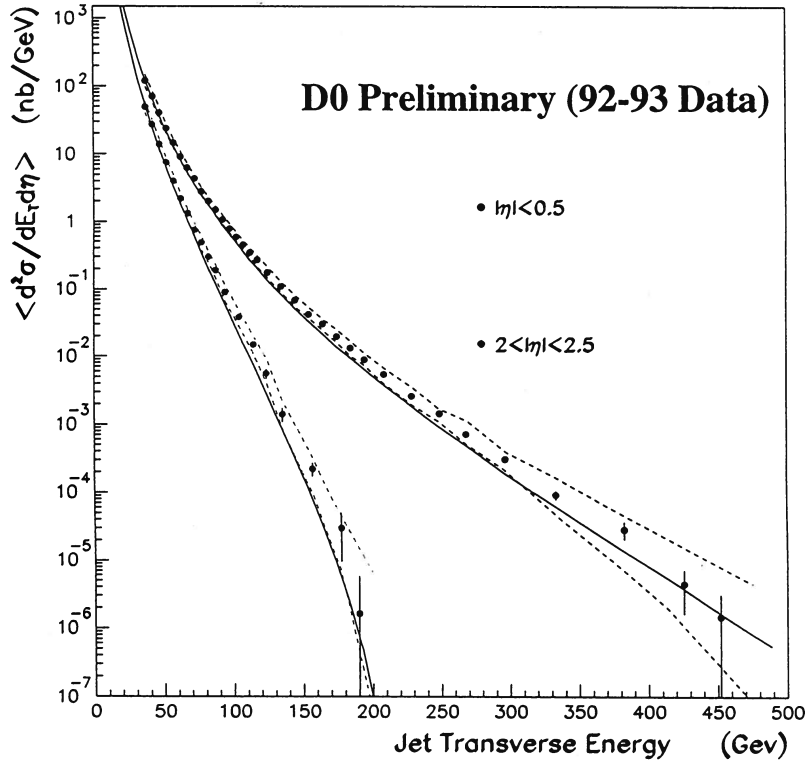}
}
\caption
{\footnotesize Inclusive jet cross-section plotted against transverse 
energy$^{\scriptsize \cite{fig8} \rm}$. \normalsize}
\label{incl.jet}
\end{figure}

Fig.~\ref{incl.jet} shows the unpolarized inclusive jet data from the D0 and CDF 
detectors$^{\scriptsize \cite{fig8} \rm}$. 
These data indicate that jets and thus probably pions, kaons 
and antiprotons could be precisely measured, with high accuracy, at the maximum 
$P_{\perp}^2$ of 54~(GeV/c)$^2$ available at 120~GeV/c and 67~(GeV/c)$^2$ at 150~GeV/c. 
Unfortunately, it is very difficult to measure the 2-spin inclusive $A_{NN}$ 
from a solid polarized proton target (PPT) due to the unpolarized protons and neutrons 
in the NH$_3$ target's beads, and in the PPT's helium coolant and bead container. 
However, it is straightforward to measure 1-spin inclusive asymmetries ($A_N$) from a 
liquid hydrogen target. 

Note that by adding two simple threshold Cherenkov counters 
to the elastic recoil spectrometer shown in Fig.~\ref{spectrom}, one could precisely 
measure inclusive cross sections, as was done in 1967-69 at 
the ZGS$^{\scriptsize \cite{incl-ZGS} \rm}$ and in 1971 at 
the ISR$^{\scriptsize \cite{incl-ISR} \rm}$. 
Since the inclusive pion, kaon and antiproton production cross sections at 
large $P_{\perp}^2$ are far larger than the elastic cross sections one could make 
rather precise $A_N$ measurements, even at $P_{\perp}^2$ of 50-70~(GeV/c)$^2$. 
Then, the prediction that for inclusive processes $A_N=0$ at large 
$P_{\perp}^2$ could be definitively tested with high precision.
 
\newpage

\vspace{0.005in}
\noindent
\bf 4. Polarized Total Cross Sections. \rm 
The Main Injector polarized beam could allow extending the 2-spin transverse proton-proton 
total cross section ${\sigma_{tot}A_{NN}}$ measurements to 120-150 GeV/c. 
One could use the traditional beam absorbtion technique with circular scintillators of 
decreasing radius followed by extrapolation of their measured rates back to zero radius. 
This simple measurement$^{\scriptsize \cite{parker} \rm}$ was the first 
polarized beam experiment when the polarized ZGS beam first operated in 1973. 
Measuring ${\sigma_{tot}}$ is far easier in fixed target experiments than in collider experiments.

\clearpage
\subsection{Updated Summary of Polarized Beam Acceleration}
\vspace{-0.1in}
\rightline{\small M.A.~Leonova, R.S.~Raymond, A.D.~Krisch \normalsize}
\vspace{0.1in}

\begin{figure}[htb]
\centerline
{
\epsfysize=3.3in \epsfbox{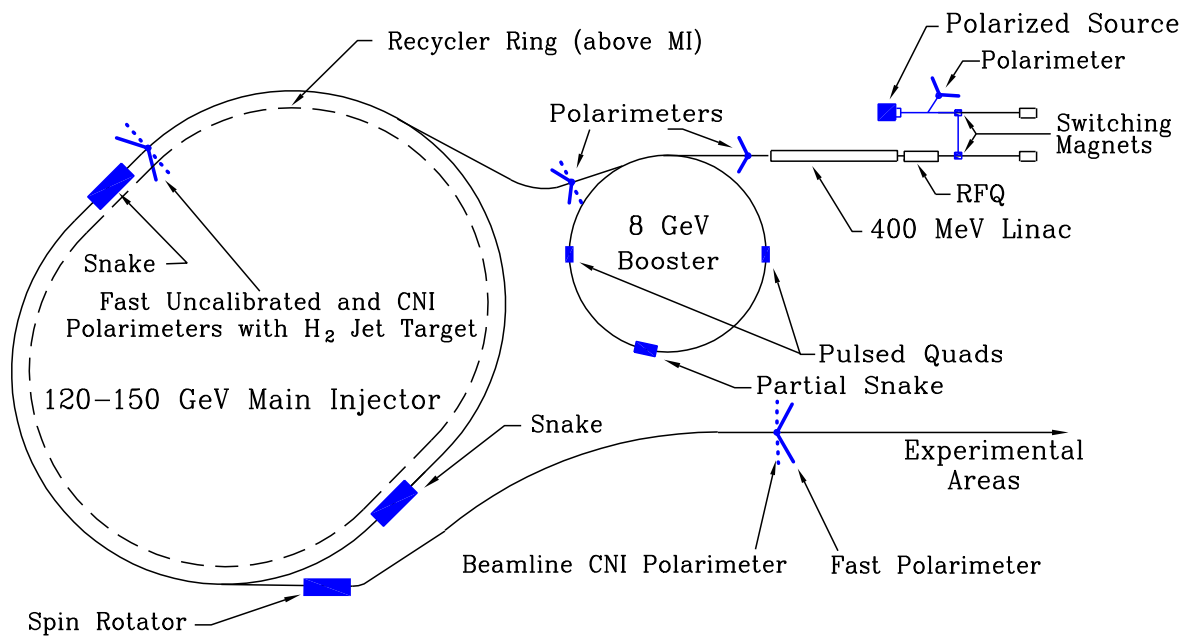}
}
\caption{\footnotesize Major items needed for polarized beam at Fermilab. \normalsize}
\label{FNAL}
\end{figure}

To accelerate polarized protons in the Main Injector, changes are needed 
in most Fermilab accelerator stages as shown in Fig.~\ref{FNAL}. 
Some of these changes were discussed in the attached 1992 Polarized Main Injector 
Report$^{\scriptsize \cite{main} \rm}$; however, much of the information is now out of date. 
The "searchable" attached 1995 Report$^{\scriptsize \cite{95rep} \rm}$, whose relevant sections 
are listed below, contains many more details, which could help one to follow this brief Updated Report.

\vspace*{0.1in}

\begin{tabular}{lll}
$\bullet$ Section 3 &p.49 & Polarized Beam Intensity\\
$\bullet$ Section 4 &p.53 & Polarized Proton Accumulation in Recycler Ring \\
$\bullet$ Section 5 &p.55 & High Intensity Polarized H$^-$ Source\\
$\bullet$ Section 6 &p.67 & RFQ for polarized H$^-$\\
$\bullet$ Section 7 &p.85 & Low Energy Beam Transport\\
$\bullet$ Section 8 &p.91 & Booster Resonance Correction\\
$\bullet$ Section 9 &p.107 & Main Injector Siberian Snakes\\
$\bullet$ Section 12 &p.147 & Polarimeters\\
$\bullet$ Section 14 &p.181 & Spin Rotation in Transport Lines \\
$\bullet$ Section 15 &p.187 & Computer Controls and Interfaces\\
\end{tabular}

\protect\vspace*{0.2in}

\noindent The major new items or changes needed are shown below and summarized in this Updated Report, where we discuss polarized beam acceleration in the current Main Injector.

\newpage
\noindent 
\bf Accelerator Modifications \rm 

\vspace{0.01in}
\noindent 
\bf 1. Polarized Ion Source: \rm 
Polarized ion sources now have intensities 
of 1.0 - 1.5$\;m$A$^{\scriptsize \cite{ABS,OPPIS} \rm}$. 
Either the former IUCF Atomic Beam type (ABS) polarized ion source (which 
is now at Dubna), or the reconstructed and improved ZGS/AGS ABS, could provide 
$\sim$1$\;m$A.

\vspace{0.01in}
\noindent
\bf 2. Beam Transport Line from 35 keV Polarized Source to RFQ: \rm 
Polarized source could share Fermilab's new RFQ with 
unpolarized sources by using new bending magnets to switch 
between sources each cycle. No depolarization if all bends are horizontal.

\vspace{0.01in}
\noindent
\bf 3. RFQ: \rm 
No depolarization in an RFQ; no changes are needed. 

\vspace{0.01in}
\noindent
\bf 4. Beam Transport Line from 750 keV RFQ to LINAC: \rm 
No changes are needed.

\vspace{0.01in}
\noindent
\bf 5. 35 keV \& 400 MeV Polarimeters: \rm 
A 35 keV RHIC-type polarimeter could monitor the source polarization during unpolarized cycles.
A 400 MeV beam-line polarimeter with a $\sim$0.5 mm carbon target could monitor polarized cycles 
with $\sim$1\% beam loss.

\vspace{0.01in}
\noindent
\bf 6. Beam Stacking and Intensity: \rm
No changes are needed.

\vspace{0.01in}
\noindent
\bf 7. 400 MeV LINAC: \rm 
There is no depolarization in LINACs; no changes are needed.

\vspace{0.01in}
\noindent
\bf 8. 400 MeV Transport Line: \rm 
Depolarization is only 0.2\%; no changes needed.

\vspace{0.01in}
\noindent
\bf 9. 8.9 GeV/c Booster Partial Siberian Snake: \rm 
A warm 4\% partial solenoidal snake, using the Booster power supply to oscillate at 15 Hz with 
a Max $\int B \cdot dl \sim 1.33$ $T \cdot m$ at 8.9 GeV/c, should overcome all 15 imperfection 
resonances (See p. 22). One could compensate small betatron-tune shift by properly ramping ring 
quadrupoles. [{\it The AGS may have a used similar magnet.}]
[{\it Weak corrector dipoles might instead overcome the rather weak imperfection resonances 
while improving the Booster beam alignment.}] 

\vspace{0.01in}
\noindent
\bf 10. 8.9 GeV/c Booster Pulsed Quadrupoles: \rm   
Two pulsed quadrupoles (3 - 10 $\mu$s risetime) should overcome 
the one fairly weak intrinsic resonance. 

\vspace{0.01in}
\noindent
\bf 11. 8.9 GeV/c Transport Line Polarimeters: \rm
Fast relative and CNI calibrated polarimeters, sharing a Carbon or fishline target, could measure the 
relative polarization after each polarized $\sim$67 ms Booster cycle.

\vspace{0.01in}
\noindent
\bf 12. 8.9 GeV/c Transport Lines: \rm 
The Booster-RR line has intermingled horizontal and vertical bends; rotator NEEDS MORE STUDY.
No depolarization in RR-MI line.

\vspace{0.01in}
\noindent
\bf 13. 8.9 GeV/c Recycler Ring: \rm  Operating point far from any resonance (See p. 22).

\vspace{0.01in}
\noindent
\bf 14. 120-150 GeV/c Main Injector Siberian Snakes: \rm 
Two superconducting Siberian snakes in Main Injector should maintain 
$\sim$95\% of injected polarization. 
The snakes must be on opposite sides of MI with orthogonal spin rotation axes. 
The snake orbit excursions must fit inside the snake magnets' ID at 8.9 GeV/c injection. 

\vspace{0.01in}
\noindent
\bf 15. 120-150 GeV/c Polarimeters: \rm
Relative and CNI polarimeters in the 120-150 GeV/c transport line sharing a 
$\sim$0.3 mm carbon or fishline target could measure the beam polarization during polarized 
cycles with $\sim$3\% beam loss. The fast polarimeter could be calibrated against the CNI 
polarimeter, and/or the Polarized Proton Target by measuring simultaneously elastic $A_N$ 
from the beam and target (Eq. 1.1). 
[\it Internal polarimeters may be possible with a very-fast-pulsed-valve hydrogen jet target.\rm]

\vspace{0.01in}
\noindent
\bf 16. 120-150 GeV/c Transport Line Spin Rotators: \rm 
A superconducting 60$^{\circ}$ spin rotator may be needed to correct 
for the spin rotation due to the intermingled horizontal and vertical 
bends in the transfer line. NEEDS MORE STUDY.

\vspace{0.01in}
\noindent
\bf 17. Computer Controls and Interfaces: \rm 
Controls for all polarized beam hardware must be interfaced with main accelerator control computer. 

%\newpage
\vspace{0.3in}
\noindent
\bf Procedure for accelerating polarized protons \rm

\vspace{0.05in}
\noindent
SeaQuest might prefer two 3-second or three 2-second polarized cycles per minute. 
However, it might be most practical to switch from unpolarized to polarized cycles 
for one minute once every ten minutes. 
This would also give the polarized beam 10\% of the beam-time and the 
unpolarized beam 90\%. Its most important advantage over switching once per minute, 
is that it would reduce the switching frequency tenfold and allow far slower 
switching times for the switching magnets before the RFQ and most importantly
much slower time for switching the polarimeter targets in and out of the beams.
Going from 50~ms to perhaps 1~s would significantly increase the switching 
hardware's lifetime and reduce its cost. It would also reduce the targets' 
oscillations after each switch, which was a significant problem in recent 
experiments at COSY$^{\scriptsize \cite{cosy} \rm}$. 
The 2~seconds of switching time could be charged to the polarized beam time. 

\vspace{0.05in}
\noindent
The following should be done to tune the polarized beam (once after each shut-down): 

\newcounter{operstep}
\begin{list} {\alph{operstep}.}{\usecounter{operstep}}

\vspace*{-2ex}
\item 
Turn on the polarized H$^{-}$ ion source 
and measure the polarization at the 35 keV Lamb-Shift Polarimeter. 
Tune the polarized source to maximize the polarization.

\vspace*{-2ex}
\item 
Adjust the switching magnet to inject polarized rather than unpolarized H$^{-}$ 
ions into the main low energy beam transport (LEBT) line to the RFQ and LINAC. 

\vspace*{-2ex}
\item 
Measure the polarization at 400 MeV with a carbon-target polarimeter. 

\vspace*{-2ex}
\item 
Turn on the Booster partial snake and adjust its ramp and the timing of 
the pulsed quadrupoles to maximize the polarization measured by the 8.9 GeV/c 
transport line polarimeters. 

\vspace*{-2ex}
\item 
Turn on the two Main Injector snakes and measure the polarization in the 
two Main Injector internal (or possibly extracted beam) polarimeters. 
Adjust the snake currents together to maximize the polarization. 
[The two cold superconducting Main Injector Snakes stay ON during both 
polarized and unpolarized cycles.]

\vspace*{-2ex}
\item 
Adjust for slow extraction into the spin experimental area rather than 
fast extraction into the neutrino production line. 

\vspace*{-2ex}
\item 
Adjust the spin rotator in the 120-150 GeV transport line to give 
the proper spin direction in the spin experimental area. 
Measure the extracted polarization using a beam line polarimeter. 

\vspace*{-2ex}
\item 
Record the magnets' settings into a separate file for a polarized beam cycle..
\end{list}

\noindent
The following would be done during operation for switching between polarized 
and unpolarized cycles:
\newcounter{operstep2}
\begin{list} {\alph{operstep2}.}{\usecounter{operstep2}}

\vspace*{-2ex}
\item
Adjust the switching magnet to inject polarized %rather than unpolarized 
ions into the main LEBT line to the RFQ.

\vspace*{-2ex}
\item 
Load the cycle settings for a polarized beam (with a snake ramp and quadrupole pulse 
in the Booster, and a flat-top in the Main Injector with slow extraction into 
the spin experimental area).

\vspace*{-2ex}
\item 
Set 400 MeV and 8.9 GeV polarimeter targets to switch into beam only on polarized cycles, if needed.

\vspace*{-2ex}
\item 
Adjust the switching magnet to inject unpolarized %rather than polarized 
ions into the main LEBT line to the RFQ.

\vspace*{-2ex}
\item 
Load the cycle settings for an unpolarized beam. 
%(no snake ramp and quadrupole pulse in the Booster, and fast extraction from 
%the Main Injector into the neutrino production line).
\end{list}

\newpage
\subsection{Summary of Needed Polarized Hardware}

\begin{enumerate}
\item
\bf Polarized Ion Source \rm \\
The source should produce a high intensity $H_{\uparrow}^{-}$ ion beam using 
Belov-type$^{\scriptsize \cite{ABS} \rm}$ Deuterium charge exchange 
from a ground state type $H_{\uparrow}^{\circ}$ atomic beam stage. \\
Intensity: 1.0 mA \\
Pulse Length: $40-100\;\mu$s \\
Polarization: more than 75\% \\ 
Emittance: 1.5 $\pi$ mm-mrad \\
Output Energy: 35 keV \\
Pulse Frequency: 15 Hz \\
Production Time: $12-24$ months \\
Remarks: This is a specialized high maintenance device; Fermilab staff should be 
integrated early into the $H_{\uparrow}^{-}$ source program. \\
Estimated Cost: $\sim$\$600,000 \\

\vspace{-0.2in}
\item
\bf Beam Transport line from 35 keV polarized source to the RFQ.\rm \\
Normalized Emittance: 0.3 $\pi$ mm-mrad \\
Vacuum: $\sim$ $10^{-7}$ Torr \\
Hardware: 	Vacuum Pipe and Vacuum Pumps \\
		Focusing Quadrupoles and Einzel lenses \\
		Bunchers (1 or 2) \\
		Switching Magnet (1-3 Hz between $H_{\uparrow}^{-}$ line and unpolarized lines) \\ 
Building Modification Time: $\sim$3 months; Estimated Cost: $\sim$\$100,000 \\
Production Time: $\sim$12 months \\ Estimated Cost: $\sim$\$100,000 \\  

\vspace{-0.2in}
\item
\bf RFQ \rm \\
The radio frequency quadrupole preaccelerator (with power supply) should accelerate 
35 keV ions to the LINAC acceptance energy. \\
Energy: 35 keV to 750 keV \\
Frequency: 201.25 MHz \\
Ion Type: polarized $H^{-}$ \\
Transmission Efficiency: $\sim$98\%  \\
Maximum Current: 50 mA \\
Cavity length: 163 cm \\
Minimum radius:  2.6 mm \\
Normalized Emmittance: 0.3 $\pi$ mm-mrad \\
Intervane RF Voltage: 67 kV \\
RF Power:  100 kW \\
Production Time: $\sim$12 or 0 months \\
Estimated Cost: $\sim$\$400,000 or \$0 \\

\vspace{-0.2in}
\item
\bf Beam Transport line from 750 keV RFQ to LINAC.\rm \\
Production Time: No change needed. \\
Estimated Cost:  \$0 \\

\newpage

\vspace{-0.35in}
\item
\bf 35 keV and 400 MeV Polarimeters \rm \\
These polarimeters could measure the transport-line polarization before the RFQ and after the LINAC. \\
Energy and Type: \\
35 keV: Lamb-shift (kills beam; use during unpolarized cycles); \\
400 MeV: p + C $\rightarrow$ p + C (kills $\sim$1\% of beam) \\
Detectors: \\
35 keV: Quench in strong E-field; measure Lyman-$\alpha$ photons; \\ 
400 MeV: Scintillators \\
Polarization Measurement Accuracy: \\
35 keV: $\sim$2\% in $\sim$ 10 sec; \\ 
400 MeV: $\sim$2\% in $\sim$1 min \\
Production Time: $\sim$12 months \\
Estimated Costs: 35 keV: $\sim$\$ 100,000; 400 MeV: $\sim$\$100,000 

\vspace{-0.1in}\item

\bf Beam Stacking and Intensity \rm \\
One could use something like the unpolarized stacking procedure for polarized ions: 40 $\mu$s pulses 
at 15 Hz repetition rate going into the LINAC and injected into the Booster for 12 turns; 
then 6 Booster pulses injected into the Recycler Ring followed by 6 more pulses using ''slip-stacking''; 
then injection into the Main Injector ring. \\
Energy: 400 MeV injection into Booster;   8.9 GeV/c stacking in RR and MI. \\
Polarized Booster Pulse Intensity Estimate: \\
\hspace*{0.05in}(46-turn-injection) 
1.0 mA $\times$ {\bf 100} $\mu$s $\times$ $6.24\cdot10^{18}$ protons/C = $6.2\cdot10^{11}$ protons; \\
\hspace*{0.05in}(18-turn-injection) 
1.0 mA $\times$ {\bf 40} $\mu$s $\times$ $6.24\cdot10^{18}$ protons/C = $2.5\cdot10^{11}$ protons; \\
\hspace*{0.05in}(12-turn-injection) 
1.0 mA $\times$ {\bf 26} $\mu$s $\times$ $6.24\cdot10^{18}$ protons/C = $1.66\cdot10^{11}$ protons. \\
MI Pulse Intensity Estimate ($2\times 6 = 12$ Booster Pulses and $\sim$95\% transf. effic.): \\
\hspace*{0.05in} with {\bf 100} $\mu$s source pulse: $7 \cdot 10^{12}$ protons;\\
\hspace*{0.05in} with {\bf 40} $\mu$s source pulse: $2.8 \cdot 10^{12}$ protons;\\
\hspace*{0.05in} with {\bf 26} $\mu$s source pulse: $1.9 \cdot 10^{12}$ protons.\\
With 10\% polarized-beam-time, one could optimize the instantaneous and 
average intensities by varying the polarized pulses' duration, frequency, and sequencing:

\vspace{-0.1in}
\newcounter{operstep3}
\begin{list} {\alph{operstep3}.}{\usecounter{operstep3}}

\vspace{-0.1in}
\item
with {\bf 100} $\mu$s source pulse: nineteen 3-sec pulses every $10^{th}$ minute 
with flat-top of $3 - 1.5 = 1.5$ sec and slow extraction 
time of 1.5 sec giving instantaneous intensity of {\bf 4.7}$\cdot 10^{12}$ p/sec and 
average intensity {\bf 13}$\cdot 10^{12}$ p/min.

\vspace{-0.1in}
\item
with {\bf 26} $\mu$s source pulse: two 3-sec pulses every minute 
with flat-top of \\ $3 - 1.334 = 1.666$ sec and slow extraction 
time of 1.66 sec giving instantaneous intensity of {\bf 1.1}$\cdot 10^{12}$ p/sec and 
average intensity {\bf 3.8}$\cdot 10^{12}$ p/min.

\vspace{-0.1in}
\item 
with {\bf 26} $\mu$s source pulse: three 2-sec pulses every minute 
with flat-top of $2 - 1.334 = 0.666$ sec and slow extraction 
time of 0.66 sec giving instantaneous intensity of {\bf 2.9}$\cdot 10^{12}$ p/sec and 
average intensity {\bf 5.7}$\cdot 10^{12}$ p/min. 

\end{list}
\vspace*{-2ex}
No changes needed. Estimated Cost: $\sim$\$0
\vspace{-0.1in}
\item
\bf 400 MeV LINAC \rm \\
There is no depolarization in LINACs; no changes needed. Estimated Cost: $\sim$\$0

\vspace{-0.1in}
\item
\bf 400 MeV Transport Line \rm \\
Depolarization is only 0.2\%; no changes needed.
Estimated Cost: $\sim$\$0

\newpage

\item
\bf 8.9 GeV/c Booster Partial Siberian Snake \rm  \\
A ramped warm solenoid 4\% partial snake should overcome all 15 imperfection depolarizing 
resonances at $G\gamma$ = 3 to 17 (See p.22). Try to run at $\nu_y$ = 6.7. \\
Energy/Momentum: 400 MeV to 8.9 GeV/c \\
Magnetic Field Integral: sine wave with amplitude increasing from 0.14 to 1.33~T$\cdot$m \\
Ampere Turns: 10$^6$ maximum \\
Maximum Current Density: 823 A/cm$^2$ \\
Coil Dimensions: Inner Rad: 4.6 cm; Outer Rad: 16.3 cm; Length: 156 cm \\
Wire Dimensions:  12 mm $\times$ 12 mm with 5 mm diam water path \\
\hspace*{1.3in} 1080 turns (9 layers $\times$ 20 turns) \\
 Solenoid Parameters: L $\sim$20 mH; R $\sim$110 m$\Omega$ \\
 Power Supply:	Booster Power supply provides current ramped at 15 Hz into a special 
 solenoid circuit with C $\sim$8.5 mF; \\
 Peak Voltage $\sim$1 kV; Average Power $\sim$45 kW \\
 Production Time: $\sim$12 months [{\it The AGS may have a used similar magnet.}]\\
 Estimated Cost: $\sim$\$200,000 

\vspace{-0.1in}
\item
\bf 8.9 GeV/c Booster Pulsed Quadrupoles \rm \\
Two pulsed quadrupoles (with ceramic vacuum chambers and a power supply) could jump 
the Booster's intrinsic depolarizing resonance ($G\gamma = 0 + \nu_y$) near $\gamma = 3.79$. 
One could use or modify 2 of the \bf 15 \rm AGS Pulsed Quadrupoles (built by Michigan) and possibly 
1 or 2 of their power supplies (built by Brookhaven). \\
 Energy: T = 2.56 GeV (P = 3.36 GeV/c) \\
 Risetime: 3 $\mu$s (300 $\mu$s falltime) \\
 Tune shift: $\nu_y$ = 0.2 \\
 Field Gradient: 1.38 $T \cdot m^{-1}$ \\
 Geometry: 3.5 cm inside radius $\times$ 50 cm long \\
 Inductance: 5 $\mu$H \\
 Power Supply: 4.3 kV, 1.3 kA for both quadrupoles \\
 Production Time: $\sim$12 months \\
 Estimated Cost: $\sim$\$100,000 

\vspace{-0.1in}
\item
\bf 8.9 GeV/c Transport Line Polarimeter \rm \\
Fast relative and calibrated CNI polarimeters could measure the beam polarization after each
$\sim$67 ms Booster acceleration cycle using p-Carbon and p-p elastic and quasielastic 
asymmetries and CNI asymmetries, respectively (see Item 15). \\
Energy/Momentum: 8.9 GeV/c \\
Target: Probably moving CH$_2$ fishline or carbon fiber\\
Detector: Scintillators \\
Polarization Measurement Accuracy: $\sim$3\% in 4 mins \\
Production Time: $\sim$12 months \\
Estimated Cost: $\sim$ \$200,000

\vspace{-0.1in}
\item
\bf 8.9 GeV/c Transport Lines \rm NEEDS MORE STUDY \\
No depolarization in RR-Main Injector line. Booster-RR line has intermingled horizontal and 
vertical bends; needs spin Rotator. Estimated Cost: $\sim$\$100,000

\newpage

\vspace{-0.1in}
\item
\bf 8.9 GeV/c Recycler Ring \rm \\
No hardware changes are needed, but one should change the RR's $\nu_y$ to better avoid the 
$G\gamma=24-\nu_y$ intrinsic resonance, which is near the RR's fixed $\gamma$ of 9.536. 
The nearby $G\gamma=17$ imperfection resonance at $\gamma= 9.483$ may be a problem that NEEDS MORE STUDY.
Estimated Cost:  \$0

\begin{table}[h!]
\vspace{-0.1in}
\caption{\footnotesize Fermilab Booster depolarizing resonances.$^{\scriptsize \cite{95rep} \rm}$ \normalsize}
\vspace{-0.1in}
\begin{center}
\begin{tabular}{lcccl} 
\hline
 $\nu_s$   &  $\gamma$ & T (GeV) & $\epsilon$ \\ \hline
  3  &  \makebox[1in]{1.673}  &  0.631 &    0.00010   \\
  4  &  2.231  &  \makebox[1in]{1.155} &    0.00014   \\
  5  &  2.789  &  1.678 & \makebox[1in]{0.00026}  \\
  6  &  3.347  &  2.201 & 0.00110   \\
  7  &  3.905  &  2.725 & 0.00180   \\
  8  &  4.462  &  3.247 & 0.00048   \\
  9  &  5.020  &  3.771 & 0.00005   \\
  10  & 5.578  &  4.294 & 0.00013   \\
  11  &  6.136 &  4.818 & 0.00027   \\
  12  &  6.694 &  5.341 & 0.00022   \\
  13  &  7.251 &  5.863 & 0.00040   \\
  14  &  7.809 &  6.387 & 0.00042  \\
  15  &  8.367 &  6.910 & 0.00030   \\
  16  &  8.925 &  7.434 & 0.00206   \\
  17  &  9.483 &  7.957 & 0.01010   \\ 
\hline
  K=0 + $\nu_y$ & 3.781 & 2.549 & 0.0132 \\
  K=24 - $\nu_y$ & 9.611 & 8.064 & 0.0474 \\ \hline
\end{tabular}
\end{center}
\label{boost-res}
\end{table}
\begin{figure}[h!]
\vspace{-0.3in}
\centerline
{
\epsfysize=3in \epsfbox{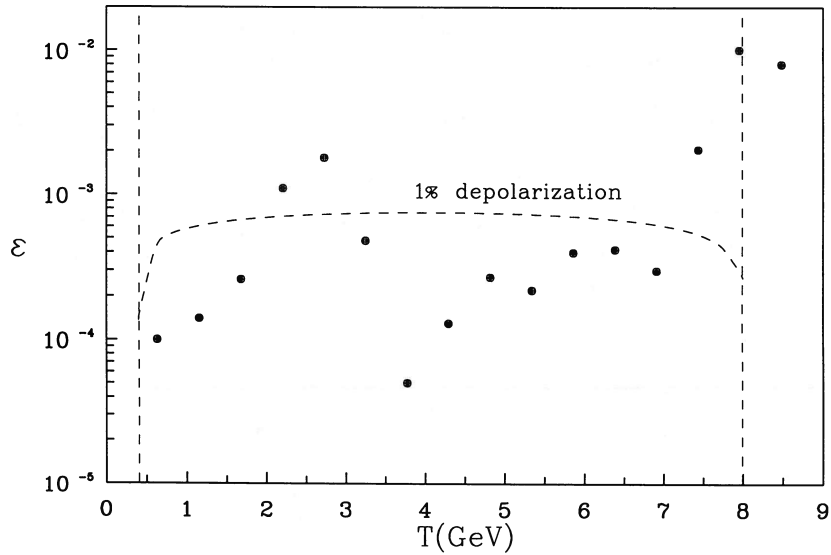}
}
\vspace{-0.1in}
\caption{\footnotesize Booster imperfection depolarizing resonance
strengths and 1\% depolarization line.$^{\scriptsize \cite{95rep} \rm}$\normalsize}
\label{boost-imp-res}
\end{figure}

\newpage
\item
\bf 120-150 GeV/c Main Injector Superconducting Siberian Snakes \rm \\
Two Siberian snakes$^{\scriptsize \cite{derb-kondr,SSC,iucf-snake} \rm}$ 
each containing 4 superconducting transverse DC helical dipole 
magnets could overcome all depolarizing resonances in the Main Injector by 
rotating the spin by 180$^{\circ}$ about a horizontal 45$^{\circ}$ axis. The two 
snakes must be placed on exactly opposite sides of the Main Injector 
ring, probably in the MI-30 and MI-60 straight sections. Moreover, to overcome strong 
depolarizing resonances the spin rotation axes of the two snakes must be orthogonal; 
for example, their axes could be $+ 45^{\circ}$ and $- 45^{\circ}$ from longitudinal.
Note that since they are superconducting, one must certify the Main Injector tunnel 
for cryogenic liquids.

We have come up with 2 different snake designs which are described below. Both are 
based on the clever and efficient RHIC 4-helical dipole 
design$^{\scriptsize \cite{snakes-RHIC,snakes-AGS} \rm}$, 
where all 4 helices are of equal lengths, with the inner pair of helices at equal 
high B-fields and the outer pair at equal lower B-fields. Moreover, each of 
the 4 dipoles has the same 360$^{\circ}$ helical rotation. 
Our new design$^{\scriptsize \cite{snakes-Mich} \rm}$ is based on 
modifying the RHIC design by making the inner pair shorter than the outer pair, so that 
the up and down vertical beam excursions are exactly equal. This minimizes the inside 
diameter of the snakes and thus significantly reduces their cost. These two new snake 
designs seem quite interesting. The snakes are fairly short with rather small 
orbit excursions, as discussed below. The 6''\,ID\,/\,B$_{max}=5$\,T snakes would require 
less time and R\&D; they have the same ID as the Main Injector; thus, they seem best from 
some points of view. However, the 4''\,ID\,/\,B$_{max}=8$\,T snakes would be 
shorter, and space for a snake in MI-60 may be an issue. Moreover, it could serve as an 
inexpensive pilot project for using superconducting NiSn magnets in high energy 
accelerators with a factor of $\sim$2000 less NiSn cable 
than the 27~km LHC. 
\begin{figure}[b!]
\centerline{
\epsfysize=3.3in \epsfbox{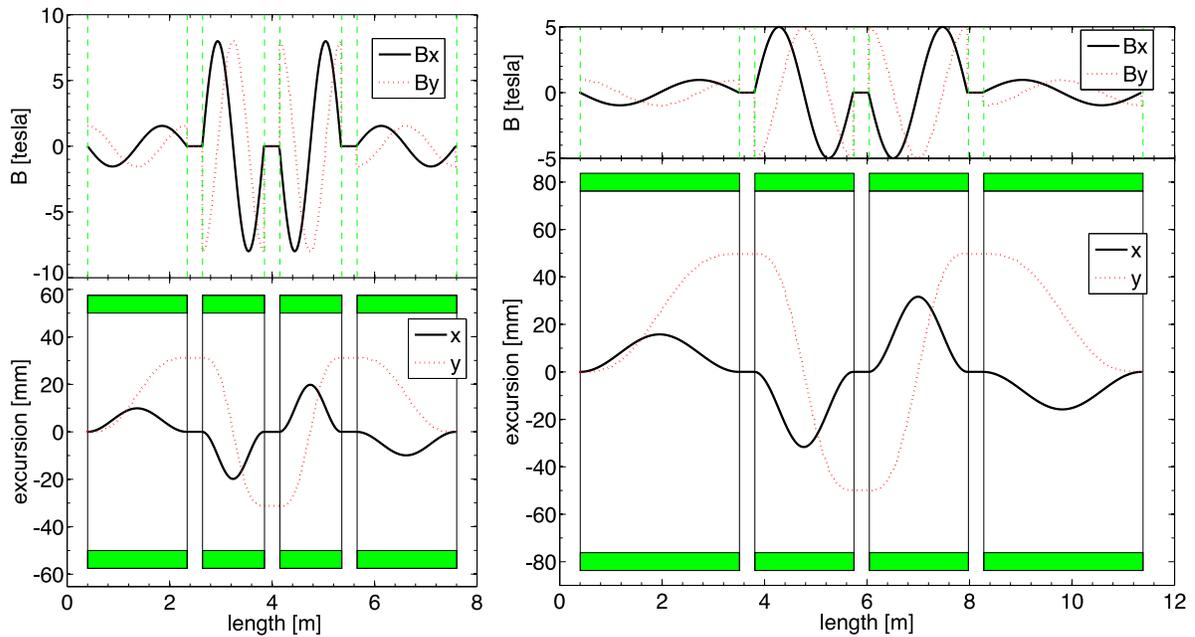}
}
\vspace{-0.05in}
\caption{\footnotesize Snakes with (LEFT) 4'' ID and B$_{max}=8$ T 
and (RIGHT) 6'' ID and B$_{max}=5$ T. \normalsize}
\label{s-4and6in}
\end{figure}
\begin{center}
\begin{table} [htb]
\vspace{-0.2in}
\center \bf Two Possible Siberian Snake Designs \rm \\
\begin{tabular} {|l|c|c|}
\hline
 Description        & 6'' ID and B$_{max}=5$ T & 4'' ID and B$_{max}=8$ T \\
\hline
 Spin Rotation Axis         & 45$^{\circ}$                & 45$^{\circ}$  \\
 Number of Helical Dipoles  & 4                           & 4     \\
 Helical Dipole's Lengths   & 3.108 m (2) \& 1.934 m (2)  & 1.943 m (2) \& 1.548 m (2) \\
 Total Snake Length         & 11.6 m                      & 7.8 m       \\
 B$_{max}$                  & 5 T                         & 8 T  \\
 Max Hor Excursion          & 31.7 mm                     & 19.8 mm \\
 Max Vert Excursion         & 49.8 mm                     & 31.2 mm \\
 Magnet Aperture            & 6'' ID (152 mm ID)          & 4'' ID (100 mm ID) \\
\hline
\end{tabular}
\vspace{-0.20in}
\end{table}
\end{center}
Momentum: 8.9 GeV/c to 120-150 GeV/c \\
Production Time: 24 months \\
Estimated Cost: \$600,000 (based on RHIC snake cost)

\vspace{-0.10in}
\item
\bf 120-150 GeV/c Transport-Line \& Possibly Internal Polarimeters \rm \\
Two transport-line polarimeters pointed at one target could measure the 
beam polarization after each polarized MI acceleration cycle. 
One polarimeter should be fast but only relatively calibrated for beam tuning. 
The other polarimeter could be slow but should be absolutely calibrated. 
A fast relative and a CNI polarimeter in the 120-150 GeV/c transport line sharing a 
$\sim$0.2 mm carbon or fishline target could measure the beam polarization after each 
polarized cycle with $\sim$2\% beam loss. 
The fast polarimeter could be calibrated against the CNI polarimeter, 
[{\it and/or the Polarized Proton Target by measuring simultaneously the elastic $A_N$ 
from the beam and target (See Eq. 1.1)}]. 
The Coulomb Nuclear Interference (CNI) polarimeter would measure the left-right asymmetry 
in proton-proton elastic scattering in the CNI region$^{\scriptsize \cite{schwinger} \rm}$ 
[P$_{\perp}^2$ $\sim$0.003 (GeV/c)$^2$] using very small recoil hodoscopes very 
near~$90^{\circ}_{lab}$.  
The Fast polarimeter could use 2 small scintillator arms to detect p-Carbon and p-p elastic and quasielastic scattering. 
[{\it Internal fast and CNI polarimeters may be possible with very-fast-valve-pulsed hydrogen jet target.}]  \\ 
Production Time: $\sim$12 months \\
Estimated Cost: $\sim$\$200,000 transport line + $\sim$\$200,000 internal

\vspace{-0.10in}
\item
\bf 120-150 GeV/c Transport Line Spin Rotator \rm NEEDS MORE STUDY \\ 
There is significant spin rotation (perhaps 60$^{\circ}$)in the MI to experimental areas transfer lines. 
This  could be compensated by a cold helical spin rotator in the 120-150~GeV/c transfer line.
The  rotator would be somewhat similar to the eight 90$^{\circ}$ rotators in RHIC. \\
Production Time: $\sim$24 months \\
Estimated Cost: $\sim$\$300,000 (based on RHIC rotator cost)

\vspace{-0.10in}
\item
\bf Computer Controls and Interfaces \rm \\
Controls for all polarized beam hardware must be conveniently and reliably interfaced with 
the main accelerator control computer. \\
Production Time: $\sim$12 months \\
Estimated Cost: $\sim$\$200,000 \\

\end{enumerate}

\clearpage
\subsection{Hardware Instalation and Schedule}

\begin{figure}[h!]
\centerline{
\epsfysize=8.3in \epsfbox{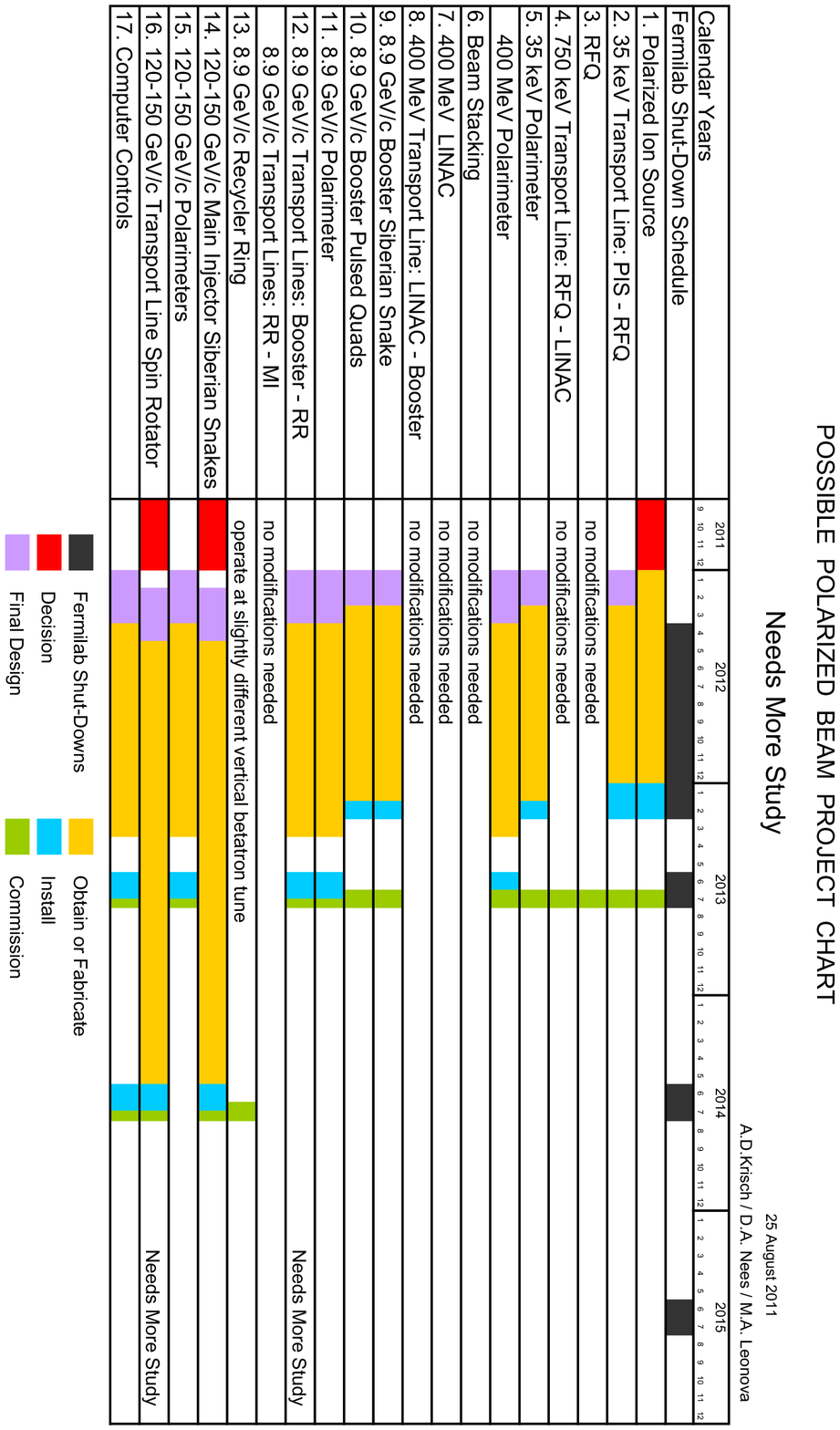}
}
\caption{\footnotesize Possible Project Chart. \normalsize}
\label{com1}
\end{figure}

The below schedule assumes that: 
\begin{enumerate}
\vspace*{-1.5ex}
\item 
the funding decision for this polarized beam project is made by December 2011; 
\vspace*{-1.5ex}
\item  
the IUCF polarized source now at Dubna will be available; 
\vspace*{-1.5ex}
\item  
Brookhaven will help to build the superconducting snakes and rotator; 
\vspace*{-1.5ex}
\item  
the switching magnets and vacuum pipes for the polarized source 
will be installed along with the RFQ; 
\vspace*{-1.5ex}
\item 
the used AGS partial snake and pulsed quadrupole are available; 
\vspace*{-1.5ex}
\item 
the ion source area (or some nearby area) is accessible during MI running. 
\vspace*{-1.5ex}
\end{enumerate}
Then one could install some of the necessary hardware during the presently 
planned FY 2012 Main Injector upgrade period.  
As shown in Fig.~\ref{com1}, about 2 years seems an appropriate time for the 
polarized beam engineering design and fabrication. \\ 
A possible commissioning sequence is:
\newcounter{abcd}
\begin{list} {\alph{abcd}.}{\usecounter{abcd}}
\vspace*{-1ex}
\item 
The polarized ion source is a critical path item. It should be obtained promptly if
Dubna agrees to sell or lease the former IUCF source. Then it could be installed and commissioned 
during the long shutdown. [{\it If not, the reconstructed ZGS/AGS polarized source might be ready 
by late 2013. Then it could be installed and commissioned during the 2014 summer shutdown.}]
\vspace*{-2ex}
\item 
The polarized source's 35 keV transport lines, and the source switching magnets should be
fabricated and installed along with the RFQ.
\vspace*{-2ex}
\item 
The 35 keV polarimeter could then be installed and commissioned along with the polarized source. 
\vspace*{-2ex}
\item 
The 400 MeV, 8.9 GeV/c and hopefully two 120-150 GeV/c polarimeters could be fabricated by early 2013. 
Then they could be installed and commissioned during the 2013 Summer shutdown. 
\vspace*{-2ex}
\item  
The Booster's partial snake and fast tune-jumping quadrupoles should be obtained promptly. They could then be installed during the long shutdown. Then there could be Booster polarized beam studies, involving injection, acceleration, and extraction during the 2013 Summer shutdown. 
\vspace*{-2ex}
\item 
The superconducting Siberian snakes and 60$^\circ$ rotator are the second critical path item. 
They should be obtained promptly if Brookhaven agrees to fabricate them. 
Then they could be installed and commissioned during the 2014 Summer shutdown.
\vspace*{-2ex}
\item 
During late 2014 polarized protons could be injected into the Main Injector for
commissioning in the 10\% of beam-time mode.  
\end{list}

\clearpage
\subsection{Commissioning} 

The Fermilab polarized beam commissioning might be somewhat similar to the 
AGS polarized beam commissioning$^{\scriptsize \cite{AGS} \rm}$, where the polarized proton
source, the 750~keV RFQ and the 200~MeV LINAC were first tuned using a
200~MeV polarimeter. The polarized 200~MeV LINAC beam was then transported to
the AGS and accelerated. In the Fermilab Booster, 
the partial snake should prevent depolarization by the imperfection 
resonances and the fast tune shifting quads should overcome the 
intrinsic depolarizing resonances.  We expect to maintain the full source 
polarization of 75\%-80\% during acceleration in the Booster and  
transport to the Main Injector.  We would then measure the 
Main Injector polarization during and after acceleration  and may 
tune its Siberian snakes if necessary.  We may
also adjust the Main Injector orbits, tunes, and 
chromaticity to maintain good emittance, intensity, and polarization.  
Polarized beam would then be extracted and transported to the experimental areas; the 
beam transport line rotators would then be tuned. This item NEEDS MORE STUDY.

\newpage

\newpage
\clearpage
\subsection{Estimated Budget}
\small
%\begin{center}
\begin{tabular}{llll}
{\bf ~Preaccelerator} &          && \$0.9M \\
  ~~Polarized H$^{-}$ ion source    && \$0.6M & \\
  ~~35 keV polarimeter             && \$0.1M & \\
  ~~RFQ and power supply (35 keV to 750 keV) && \$0.0M & \\
  ~~Beam lines, switching magnets \& vacuum system   && \$0.1M & \\
  ~~Building Modification           && \$0.1M & \\
{\bf ~400 MeV LINAC}                 &&& \$0.1M  \\
  ~~400 MeV polarimeter             && \$0.1M & \\
{\bf ~8.9 GeV/c Booster}                 &&& \$0.6M  \\
  ~~Solenoid partial Siberian snake (ramped warm) &&\$0.2M & \\
  ~~Two 3 $\mu$sec pulsed quadrupoles with power supplies && \$0.1M & \\
  ~~8.9 GeV/c polarimeter     && \$0.2M & \\
  ~~8.9 GeV/c transfer line spin rotator && \$0.1M & \\
{\bf ~Main Injector}                 &&& \$0.9M \\  
  ~~Two Helical Siberian snakes     && \$0.6M & \\
  ~~Power supplies for snakes       && \$0.1M & \\
  ~~120-150 GeV/c polarimeters (CNI \& Inclusive) && \$0.2M & \\
{\bf ~120-150 GeV/c Transfer Line}                 &&& \$0.5M \\
  ~~120-150 GeV/c polarimeters (CNI \& Inclusive) && \$0.2M & \\
  ~~120-150 GeV/c transfer line spin rotator && \$0.3M & \\
{\bf ~Miscellaneous}                 &&& \$0.2M \\
  ~~Computers, control modules, cables, and interface && \$0.2M & \\
{\bf ~Main Injector subtotal}        &&&  \$3.2M \\
{\bf ~Contingency ($\sim$25\%)}     &&&  \$0.8M \\ 
{\bf MAIN INJECTOR TOTAL}           &&& {\bf $\sim$\$4M} \\ 
\end{tabular}
%\end{center} 

\vspace*{0.2in}
\normalsize 
The estimate for the total cost of obtaining 120-150 GeV/c polarized
proton beam capability at Fermilab is given in 2012 Dollars.

\subsection{Summary}
With a 50 cm long liquid hydrogen target and 10\% of the beam time, 
the time-averaged polarized beam luminosity for the Main Injector could 
probably be about $2\cdot10^{35}$cm$^{-2}$s$^{-1}$ or higher. 
This polarized luminosity should allow precise 
measurements of spin-asym\-metries out to \mbox{$P^2_{\perp}$} of 50-70 (GeV/c)$^2$ 
for inclusive hadron production. The world's highest intensity polarized proton beam with 
a 50 cm hydrogen target would also allow precise studies of polarized Drell-Yan processes. 
With a solid polarized proton target, it could also allow high-precision 1-spin, 2-spin and 
spin-averaged studies of violent elastic proton-proton collisions out to \mbox{$P^2_{\perp}$} 
of at least 12 (GeV/c)$^2$ - a fundamental probe of the strong interaction.

The total cost of providing a 120-150~GeV/c polarized proton beam could be about 
\$4~Million and the time needed for producing the needed hardware could be
about 24~months from the time of approval and funding.

\newpage

\addcontentsline{toc}{subsection}{References}

%\begin{thebibliography}{99}
\refer
\bibitem{moss}
J.M. Moss {\it et al.}, Draft Letter of Intent, 10 May 1995.
\bibitem{ref1.1}
P.R. Cameron {\it et al.}, Phys. Rev. Rapid Comm. {\bf D32}, 3070 (1985);\\
D.G. Crabb {\it et al.}, Phys. Rev. Lett. {\bf 64}, 2627 (1990).
\bibitem{ref1.2}
%K. Heller, Proc. 7th International Symposium on High Energy Spin Physics, Protvino, 1986.
C.A. Aidala, {\it Proc. of 18$^{th}$ Int. Spin Physics Symposium}, 
AIP {\bf 1149}, 124 (2009); \\
D.G.~Crabb on behalf of SPIN08 organizing committee, CERN Courier June (2009).
\bibitem{ref1.3}
J.R. O'Fallon {\it et al.}, Phys. Rev. Lett. {\bf 39}, 733 (1977);\\
D.G. Crabb {\it et al.}, Phys. Rev. Lett. {\bf 41}, 1257 (1978).
\bibitem{fig4}
A.D. Krisch, Phys. Rev. Lett. {\bf 19}, 1149 (1967); \\
P.H.~Hansen and A.D.~Krisch, Phys. Rev. {\bf D15}, 3287 (1977);\\
A.D. Krisch, Z. Phys. {\bf C46}, S113 (1990).
\bibitem{UG}
G.E. Uhlenbeck and S. Goudsmit, Naturwiss. {\bf 13}, 953 (1925).
\bibitem{yang}
C.N. Yang, AIP Conf. Proc. {\bf 95}, 1 AIP, New York (1983).
\bibitem{wolf}
L. Wolfenstein, Phys. Rev. {\bf 75}, 1664 (1949); \\
L. Wolfenstein and J. Ashkin, Phys. Rev. {\bf 85}, 947 (1952).
\bibitem{oxley}
C.L Oxley {\it et al.}, Phys. Rev. {\bf 93}, 806 (1954).
\bibitem{fermi}
E. Fermi, Il Nuovo Cimento {\bf 93}, 11 (1954).
\bibitem{crosbie}
E.A. Crosbie {\it et al.}, Phys Rev. {\bf D23}, 600 (1981).
\bibitem{KPR}
G. Kane, J. Pumplin and W. Repko, Phys. Rev. Lett. {\bf 41}, 1689 (1978).
\bibitem{sivers1}
D.W. Sivers, Phys. Rev. {\bf D41}, 83 (1990); {\bf D43}, 261 (1991).
\bibitem{BM-dis}
D. Boer and P.J. Mulders, Phys. Rev. {\bf D57}, 5780 (1998).
\bibitem{collins1}
J.C. Collins, Nucl. Phys. {\bf B396}, 161 (1993).
\bibitem{ff}
R.D. Tangerman and P.J. Mulders, Phys Rev. {\bf D51}, 3357 (1995); 
Nucl. Phys. {\bf B461}, 197 (1996).
\bibitem{sivers2}
D.W. Sivers, Phys Rev. {\bf D74}, 094008 (2006).
\bibitem{trento}
A. Bacchetta, U. D'Alesio, M. Diehl and C.A. Miller, Phys Rev. {\bf D74}, 094008 (2006).
\bibitem{CHL}
J.C. Collins, S.F. Heppelmann and G.A. Ladinsky, Nucl. Phys. {\bf B420}, 565 (1994).
\bibitem{RS}
J.P. Ralston and D.E. Soper, Nucl. Phys. {\bf B152}, 109 (1979).
\bibitem{JJ}
R.L. Jaffe, X.-D. Ji, Phys. Rev. Lett. {\bf 67}, 552 (1991).
\bibitem{DY}
S.D. Drell and T.-M. Yan, Phys. Rev. Lett. {\bf 25}, 316 (1970).
\bibitem{Anselmino}
M. Anselmino {\it et al.}, Nucl. Phys. Proc. Suppl. {\bf 91}, 98 (2009), arXiv:0812.4366[hep-ph].
%P. Kroll and B. Pire, Z. Phys. {\bf 36}, 89 (1987). 
\bibitem{collins2}
J.C. Collins, Phys. Lett. {\bf B536}, 43 (2002), arXiv:hep-ph/0204004.
\bibitem{CY}
T.T. Chou and C.N. Yang, Phys. Lett. {\bf 135}, 175 (1984).
\bibitem{LB}
G.P. Lepage and S.J. Brodsky, Phys. Rev. {\bf D22}, 2157 (1980).
\bibitem{hyp-incl}
G. Bunce {\it et al.}, Phys. Rev. Lett. {\bf 36}, 1113 (1976); \\ 
K. Heller {\it et al.}, Phys. Rev. {\bf D16}, 2731 (1977).
\bibitem{E904}
A. Yokosawa et al.,Phys. Lett. {\bf B261}, 201 (1991); Phys. Lett. {\bf B264}, 462 (1991).
\bibitem{ZGS}
T.~Khoe {\it et al.}, Part. Accel {\bf 6}, 213 (1975).
\bibitem{AGS} 
F.Z. Khiari {\it et al.}, 
%{\it Acceleration of Polarized Protons to 22 GeV/c and measurement of spin-spin effects}, 
Phys. Rev., {\bf D39}, 45 (1989).
\bibitem{Brodsky2002} 
S.J. Brodsky {\it et al.}, Nucl. Phys. \textbf{B642}, 344 (2002), arXiv:hep-ph/0206259.
\bibitem{hermes2005} 
A. Airapetian, {\it et al.}, (HERMES Collaboration), 
Phys. Rev. Lett. \textbf{94}, 012002 (2005), \\
arXiv:hep-ex/0408013.
\bibitem{compass2005} 
V.Y. Alexakhin, {\it et al.}, (COMPASS Collaboration), 
Phys. Rev. Lett. \textbf{94}, 202002 (2005), \\
arXiv:hep-ex/0503002. 
\bibitem{VQL}
M. Anselmino {\it et al.}, EPJA. {\bf 39}, 89 (2009).
\bibitem{fig8}
H. Weerts, private communications.
\bibitem{incl-ZGS}
L.G.~Ratner {\it et al.}, Phys. Rev. Lett {\bf 18}, 1218 (1967);\\
L.G.~Ratner {\it et al.}, Phys. Rev. {\bf 166}, 1353 (1968);\\
G.J.~Marmer {\it et al.}, Phys. Rev. Lett. {\bf 23}, 1469 (1969).
\bibitem{incl-ISR}
L.G.~Ratner {\it et al.}, Phys. Rev. Lett. {\bf 27}, 68 (1971).
\bibitem{parker} 
E.F. Parker {\it et al.}, Phys. Rev. Lett. {\bf 31}, 783 (1973); \\
W. de Boer {\it et al.}, Phys. Rev. Lett. {\bf 34}, 558 (1975).
\bibitem{main}
{\em Report on Acceleration of Polarized Protons to 120 and 150 GeV in the
Fermilab Main Injector}, SPIN Collaboration, Unpublished University of
Michigan Report UM HE 92-05 (March 1992).
\bibitem{95rep}
{\em Report on Acceleration of Polarized Protons to 120~GeV and 1~TeV at 
Fermilab}, SPIN Collaboration, Unpublished University of
Michigan Report UM HE 95-09 (July 1995).
\bibitem{ABS} 
A.S. Belov {\it et al.}, private communications.
\bibitem{OPPIS} 
A.N. Zelenski {\it et al.}, SPIN 2008.
%IPAC-95 Dallas Accelerator Conference, May 1995.
\bibitem{cosy}
V.S. Morozov {\it et al.}, Phys. Rev. Lett. {\bf 103}, 144801 (2009); \\
V.S. Morozov {\it et al.}, Phys. Rev. Lett. {\bf 102}, 244801 (2009); \\
V.S. Morozov {\it et al.}, Phys. Rev. Lett. {\bf 100}, 054801 (2008).
\bibitem{derb-kondr}
Ya.S. Derbenev and A. M. Kondratenko, Sov. Phys. JETP 35, 230 (1972); \\ 
Ya.S. Derbenev {\it et al.}, Part. Accel 8, 115 (1978).
\bibitem{SSC} 
{\it Proc. of 1985 Ann Arbor Workshop on Polarized Beams at SSC}, eds. A.D.~Krisch, 
A.M.T.~Lin and O.~Chamberlain, AIP Conf. Proc. {\bf 145} (AIP, New York 1986).
\bibitem{iucf-snake}
A.D. Krisch {\it et al.}, Phys. Rev. Lett. {\bf 63}, 1137 (1989).
\bibitem{snakes-RHIC} 
M. Syphers {\it et al.}, 
%''Helical Dipole Magnets for Polarized Protons in RHIC'', 
{\it Proc. of 1977 Part. Accel. Conf.}, 3359 (1998); \\
E. Willen, {\it et al.}, 
%''Construction of Helical Magnets for RHIC'', 
{\it Proc. of 1999 Part. Accel. Conf.}, 3161 (1999); \\
E. Willen, {\it et al.}, 
%''PERFORMANCE SUMMARY OF THE HELICAL MAGNETS FOR RHIC'', 
{\it Proc. of 2003 Part. Accel. Conf.}, 164 (2003).
\bibitem{snakes-AGS}
E. Willen, {\it et al.}, 
%''SUPERCONDUCTING HELICAL SNAKE MAGNET FOR THE AGS'', 
{\it Proc. of 2005 Part. Accel. Conf.}, 2935 (2005).
\bibitem{snakes-Mich} 
M.A.~Leonova, E.D. Courant and A.D. Krisch, in preparation.
\bibitem{schwinger}
J. Schwinger Phys. Rev. {\bf 73}, 407 (1948).
\end{thebibliography}
\normalsize

\addtocontents{toc}{ 
\noindent ~~\\ \noindent \bf 2~~Appendices \rm (Under separate Cover) \\ 
\vspace*{-0.15in} \\
%\noindent 
2.1  UM HE 92-05 (1992) Report on Acceleration of Polarized Protons 
to 120 and 150~GeV in the Fermilab Main Injector. \\ 
\vspace*{-0.15in} \\
%\noindent 
2.2  UM HE 95-09 (1995) Report on Acceleration of Polarized Protons 
to 120~GeV and 1~TeV at Fermilab.}
%\section*{Appendix: Report}

\end{document}